\documentclass{vgtc}                          

\ifpdf
  \pdfoutput=1\relax                   
  \usepackage{graphicx}                
  \DeclareGraphicsExtensions{.pdf,.png,.jpg,.jpeg} 
\else
  \usepackage{graphicx}                
  \DeclareGraphicsExtensions{.eps}     
\fi%

\graphicspath{{fig/}{./}} 

\usepackage{microtype}                 
\PassOptionsToPackage{warn}{textcomp}  
\usepackage{textcomp}                  
\usepackage{mathptmx}                  
\usepackage{times}                     
\usepackage{cite}                      
\usepackage{tabu}                      
\usepackage{booktabs}                  

\usepackage{xcolor}
\usepackage{caption}
\usepackage{amsmath}
\usepackage{amssymb}
\usepackage{algorithm}
\usepackage{algorithmic}
\usepackage[export]{adjustbox}
\usepackage{graphicx}
\usepackage{booktabs}
\usepackage{colortbl}
\usepackage{enumitem}
\usepackage{verbatim}
\usepackage{wrapfig}
\usepackage{hyperref}

\newcommand{\fn}[1]{{\textbf{{#1}}}}

\onlineid{0}

\vgtccategory{Research}

\vgtcinsertpkg

\hypersetup{
pdftitle={TopicSifter: Interactive Search Space Reduction Through Targeted Topic Modeling},
pdfsubject={VAST 2019},
pdfauthor={Hannah Kim, Dongjin Choi, Barry Drake, Alex Endert, Haesun Park},
pdfkeywords={Text analytics, document retrieval, topic modeling, nonnegative matrix factorization}
}

\title{TopicSifter: Interactive Search Space Reduction \\Through Targeted Topic Modeling
}

\author{Hannah Kim\thanks{e-mail: hannahkim@gatech.edu}\\ %
     \scriptsize Georgia Institute of Technology %
\and Dongjin Choi\thanks{e-mail: jin.choi@gatech.edu}\\ %
     \scriptsize Georgia Institute of Technology %
\and Barry Drake\thanks{e-mail: barry.drake@gtri.gatech.edu}\\ %
     \scriptsize Georgia Tech Research Institute %
\and Alex Endert\thanks{e-mail: endert@gatech.edu}\\ %
     \scriptsize Georgia Institute of Technology %
\and Haesun Park\thanks{e-mail: hpark@cc.gatech.edu}\\ %
     \scriptsize Georgia Institute of Technology %
}

\teaser{
	\centering
	\includegraphics[width=0.87\textwidth]{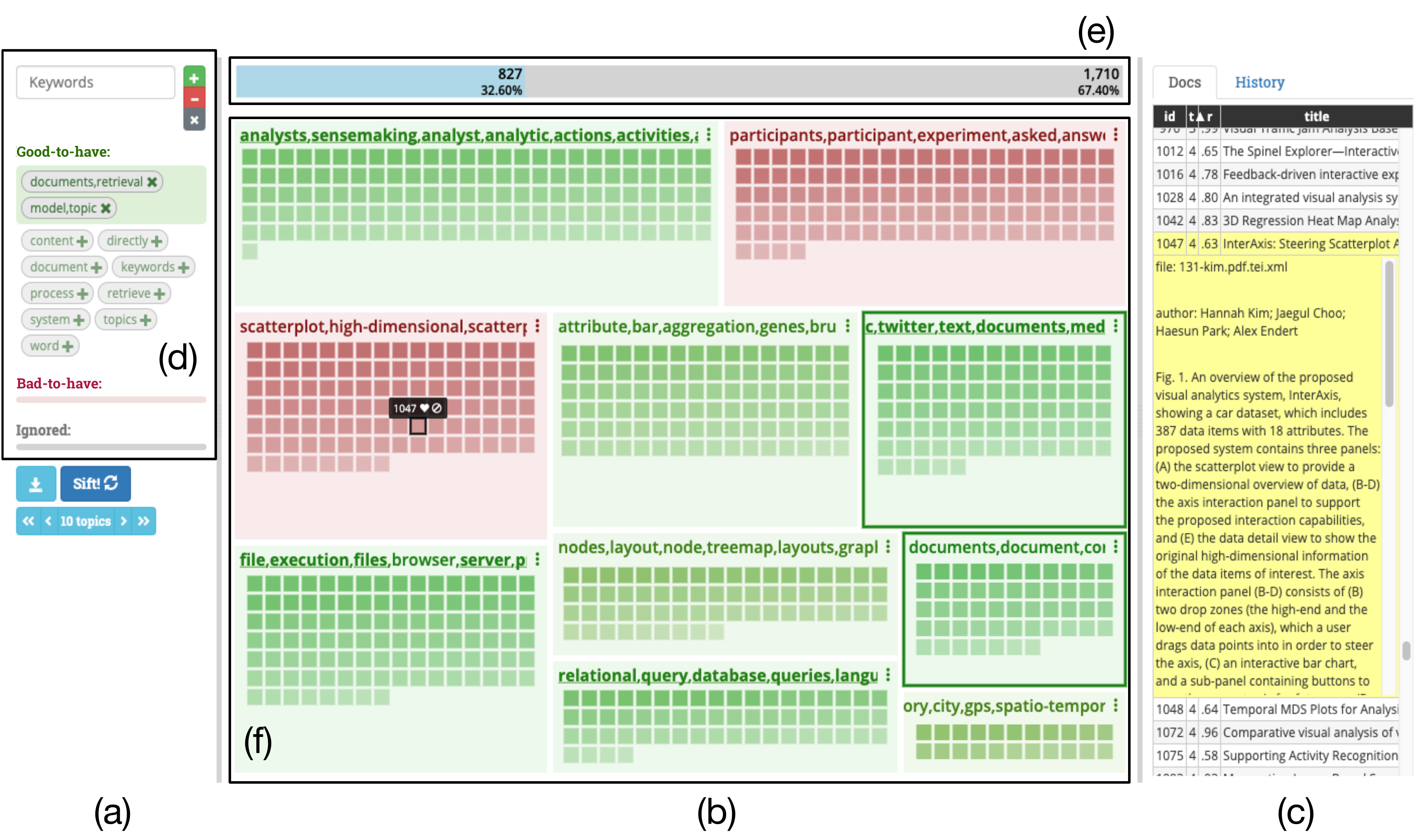}
\vspace{-0.5em}
	\caption{The TopicSifter system has (a) the control panel, (b) the main view, (c) the detail panel. 
The keyword module (d) in the control panel (a) shows the current set of good-to-have keywords, bad-to-have keywords, and stopwords and allows users to modify them. 
The system recommends additional keywords based on the current set of keywords. 
The main view (b) shows the sifting status bar (e) showing how many documents are retrieved from the total dataset and the topical overview (f) of current retrieved documents. 
The users can give positive or negative feedback on topics and documents to indicate relevancy. 
The detail panel (c) has two tab menus for showing document details and sifting history.}
	\label{fig:teaser}
}

\abstract{
Topic modeling is commonly used to analyze and understand large document collections.
However, in practice, users want to focus on specific aspects or ``targets'' rather than the entire corpus.
For example, given a large collection of documents, users may want only a smaller subset which more closely aligns with their interests, tasks, and domains.
In particular, our paper focuses on large-scale document retrieval with high recall where any missed relevant documents can be critical. 
A simple keyword matching search is generally not effective nor efficient as 1) it is difficult to find a list of keyword queries that can cover the documents of interest before exploring the dataset, 2) some documents may not contain the exact keywords of interest but may still be highly relevant, and 3) some words have multiple meanings, which would result in irrelevant documents included in the retrieved subset.
In this paper, we present TopicSifter, a visual analytics system for interactive search space reduction. 
Our system utilizes targeted topic modeling based on nonnegative matrix factorization and allows users to give relevance feedback in order to refine their target and guide the topic modeling to the most relevant results.

} 

\CCScatlist{
  \CCScatTwelve{Human-centered computing}{Visu\-al\-iza\-tion}{Visu\-al\-iza\-tion application domains}{Visual analytics};
  \CCScatTwelve{Information systems}{Information retrieval}{Users and interactive retrieval}{Search interfaces}
}

\begin{document}


\firstsection{Introduction}
\label{sec:intro}

\maketitle

As the world becomes increasingly digital and huge amounts of text data are generated every minute, it becomes more challenging to discover useful information from them for applications such as situational awareness, patient phenotype discovery, event detection~\cite{Drake2017}, or the onset of violence within a diverse population.
More often than not, topics of interests are only implicitly covered in vast amounts of text data and the relevant data items are sparse and not immediately obvious.
This scenario is more prevalent especially in large scale data analytics where the data are obtained from passive sources and not all data items are relevant to the questions at hand. 
In these cases, users want to focus on a subset of documents about specific aspects or ``targets'', rather than analyzing entire document collections~\cite{wang2016targeted}.
For example, a journalist may want to analyze social media data that are related to a specific event. 
Similarly, a marketing expert may want reviews that are relevant to certain products or brands only.
Both examples require the \textit{search space} of entire documents to be \textit{reduced} to relevant documents.

Discovering and extracting data items of relevance from a large collection of documents is a challenging and important step in text analytics.
In particular, we are interested in the \textit{high recall} retrieval problem, where any missing relevant documents are critical~\cite{li2014req}.
For instance, a legal analyst searching for relevant cases from a large legal document collection may want to collect as many documents as possible even if some of them are only slightly relevant to her targets.
Another example is a graduate student who is preparing a literature review and does not want to miss a related work.
This is different from a traditional informational retrieval problem of finding a list of $k$ results that are most relevant to a query, e.g., a student searching for the top 5 papers to learn about an unfamiliar research field. 
Our focus is on not missing relevant results in addition to high precision. 
To solve this, our goal is to retrieve documents that are relevant to targets from large scale document collections, which we will refer to as \textbf{search space reduction} throughout this paper.

\begin{wrapfigure}{r}{2.5cm}
	\centering
	\includegraphics[width=2.5cm]{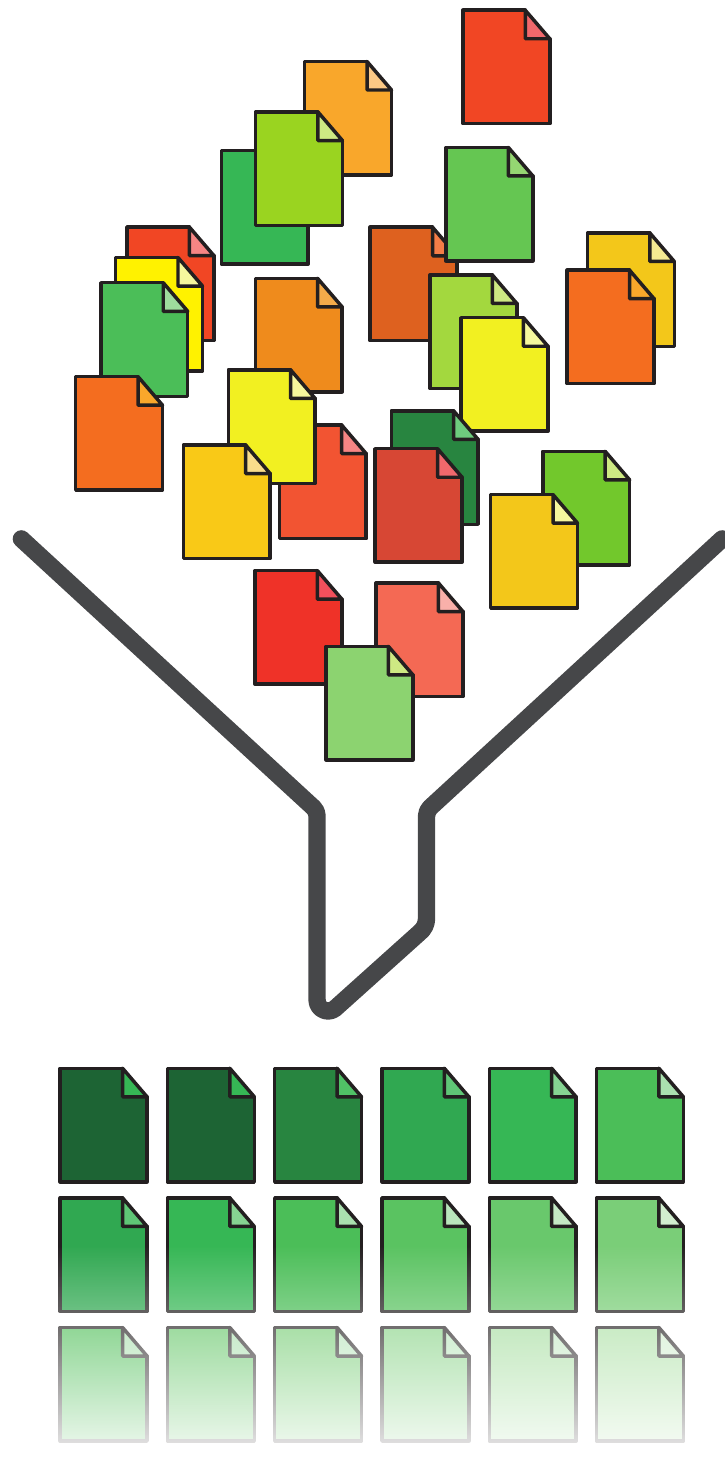}
	\caption{An illustration for search space reduction, retrieving relevant documents from large corpora with high recall.}
	\label{wrap-fig:sift}
\end{wrapfigure}

Traditional static keyword search is not suitable for our search space reduction setting.
First, it is often difficult to know or express the target aspect in advance without exploring the dataset.
Next, even when the users are familiar with their target concept, it is hard to cover all relevant keywords, which would result in false negative.
Lastly, a keyword may have multiple unrelated meanings and when they are extracted out of context, static keyword match can result in false positive. 
More advanced approaches such as query expansion and relevance feedback have been introduced in information retrieval.
These approaches expand query keywords and provide feedback on documents to update the query.
However, since they are designed for high precision problems of retrieving a number of the most relevant data items, they may not cover all relevant data items.

To this end, we take a human-in-the-loop approach and advocate interactive and exploratory retrieval.
In our framework, users explore retrieved documents, learn them, and interactively build targets, which will be used to sift through documents. 
Instead of users rating a number of retrieved documents generated by systems, our method allows the users to proactively modify target keywords and give relevance feedback.
In addition, we adopt \textit{targeted topic modeling} to support this process.
Targeted topic modeling techniques find relevant topics and disregard irrelevant aspects from document collections.
Utilizing results from targeted topic modeling, our approach allows users to discover relevant subtopics and refine the targets using topic-level relevance feedback.

In this paper, we propose a novel framework for interactive search space reduction
along with an effective visual analytics system called TopicSifter. 
TopicSifter tightly integrates the underlying computational methods and interactive visualization to support topic model exploration and targeted topic modeling. 

The primary contributions of this work include:
\begin{itemize}
\vspace{-0.5em}
\setlength{\itemsep}{-.2\baselineskip}
    \item A novel iterative and interactive technique for search space reduction through interactive target building, sifting, and targeted topic modeling.
    \item A visual analytics prototype, TopicSifter, that supports tight integration between the interactive visualization and the underlying algorithms.
    \item Experiments and use cases that illustrate the effectiveness of TopicSifter.
\end{itemize}

\section{Related Work}
\label{sec:relwork}
In this section, we discuss prior works on information retrieval and topic modeling in the context of search space reduction.

\subsection{Visualizing Search Results/Space}
Various information visualization techniques have been applied to improve user interfaces for search.
Some systems augment search result lists with additional small visualizations.
For example, TileBars~\cite{hearst1995tilebars}, INSYDER~\cite{Reiterer2005}, and HotMap~\cite{hoeber2006comparative} visualize query-document relationships as icons or glyphs alongside search results.
Another approach is to visualize search results in a spatial layout where proximity represents similarity.
Systems such as InfoSky~\cite{infosky02} and IN-SPIRE~\cite{inspire04} are examples.
FacetAtlas~\cite{facetatlas10} overlays additional heatmaps to visualize density.
ProjSnippet~\cite{gomez2014similarity} visualizes text snippets in a 2-D layout.
Many others cluster the search results and offer faceted navigation.
FacetMap~\cite{smith2006facetmap} and ResultMap~\cite{clarkson2009resultmaps} utilizes treemap-style visualizations to represent facets.
These systems may guide users well in exploring search results, but they are mostly based on static search queries.
Our system goes beyond search results exploration and offers interactive target (query) building.

\subsection{Query Expansion and Relevance Feedback}
Information retrieval is finding (unstructured) documents that satisfies an information need from large collections~\cite{manning2010introduction}.
However, users of information retrieval systems may not have a clear idea of what to search for, may not know how to construct an optimal query, or may not understand what kind of information is available~\cite{ruthven2003survey}.
To this end, various interactive methods to assist the retrieval process have been proposed.
Interactive query expansion~\cite{harman1988towards} allows the users to choose additional query terms from the suggested list of keywords.
Instead of lists, Fowler et al.~\cite{fowler1991integrating} and Hoeber et al.~\cite{hoeber2005visualization} display keyword suggestions as graphs.
Sparkler~\cite{havre2001interactive} visualize multiple query results so that users can compare and identify the best query from the expanded queries.
In our system, we suggest additional keywords for queries in terms of two categories of good-to-have keywords \textit{and bad-to-have} keywords.
Another interactive approach is relevance feedback, meaning users are asked to mark documents as relevant to steer the system to modify the original query~\cite{ruthven2003survey}.
For instance, VisIRR~\cite{visirr} allows users to rate retrieved documents on a 5-star scale.
IntentRadar~\cite{ruotsalo2013directing,ruotsalo2015interactive} models intents behind search queries and lets users give relevance feedback on the intents to interactively update them.
We adopt a similar approach to give relevance feedback to documents \textit{as well as groups of documents (topics)}.
These existing systems are designed for the traditional information retrieval setting of obtaining the most relevant data items with high precision, and thus are not well-suited for our search space reduction setting which desires high recall.
Closer to our work is ReQ-ReC~\cite{li2014req} which combines iterative query expansion and iterative classifier refinements to solve high recall retrieval problem.
A major difference is that ReQ-ReC system requires users to label given documents while our system allows the users to explore the documents and their topics and give relevance feedback if needed.

\subsection{Aspect-Specific Topic Summarization}
Although topic summarization has been studied for a long time, discovering topic summary of a specific aspect (or targets) is a relatively new research problem.
TTM~\cite{wang2016targeted} is the first work to propose the term `targeted topic modeling'.
This work proposes a probabilistic model that is a variation of latent Dirichlet allocation (LDA)~\cite{blei2003latent}.
Given a static keyword list defining a particular aspect, the model identifies topic keywords related to this aspect.
Wang et al.~\cite{wang2018disentangling} identifies a list of target words from review data and disentangles aspect words and opinion words from the list.
APSUM~\cite{rakesh2018sparse} assigns aspects to each word in a generative process.
Since the aforementioned model generates topic keywords based on a static keyword list, a dynamic model is desired.
An automatic method to generate keyword dynamically has been proposed~\cite{zheng2017semi}.
This method focuses on the on-line environment of Twitter and automatically generates keywords based on the time-evolving word graph.

\subsection{Interactive Topic Modeling}
Interactive topic models allow users to steer the topics to improve the topic modeling results.
Various topic steering interactions such as adding, editing, deleting, splitting, and merging topics have been introduced~\cite{forcespire12,ivisclustering12,utopian13,roseriver14,topiclens17,HERR201728,hltm2018,convisit2015,mcon2018}.
These interactions can be applied to refine relevant topics and remove irrelevant topics to identify targeted topics when most of the data items are relevant and only a small portion is irrelevant.
However, in our large-scale search space reduction setting, a more tailored approach is needed.
In this paper, we propose interactive targeted topic modeling to steer the topics to discover the target-relevant topics and documents.

\section{Interactive Search Space Reduction}
\label{sec:technique}
\begin{figure}[t]
	\centering
	\includegraphics[width=\linewidth]{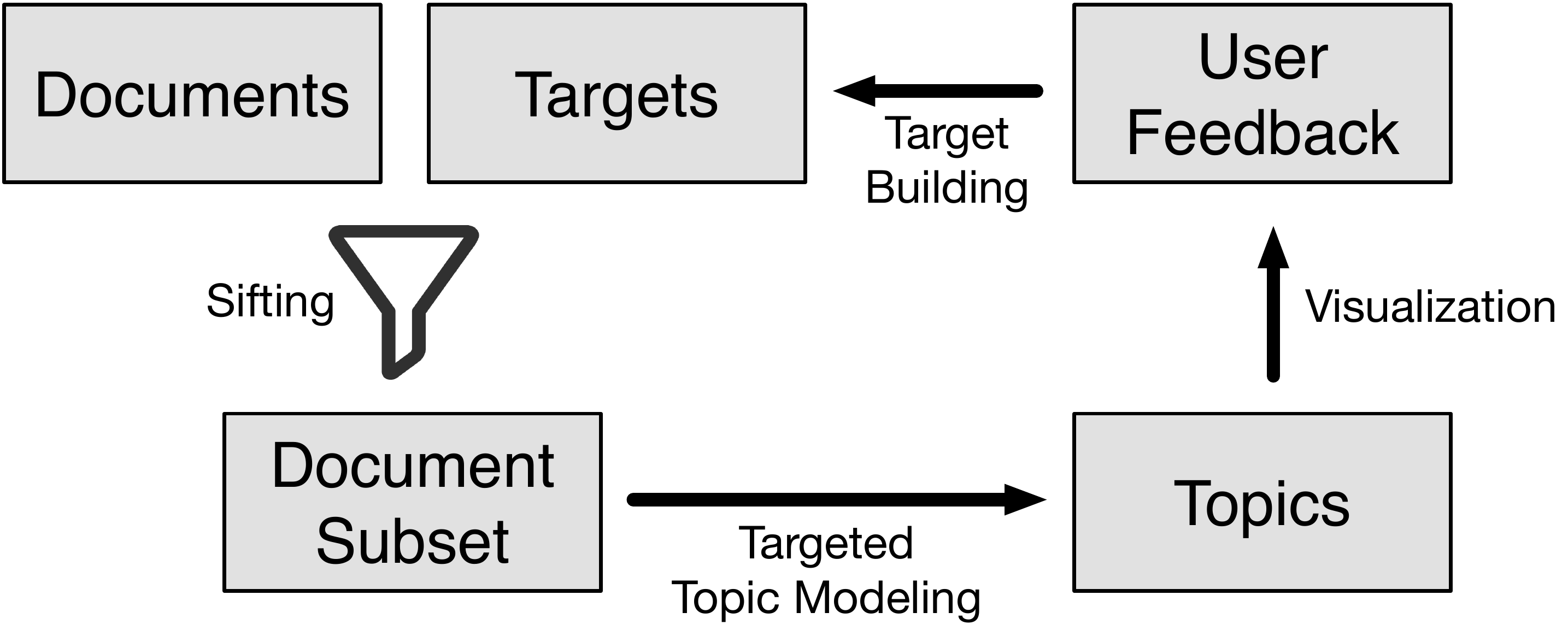}
\vspace{-1.5em}
	\caption{Our human-in-the-loop algorithm workflow for interactive search space reduction. 
A document subset is updated based on user feedback each iteration.}
	\label{fig:workflow}
\vspace{-1.5em}
\end{figure}

In many practical cases in large-scale text analyses, users have specific aspects they want to focus on, which we will refer to as targets.
Although there are many tools available with powerful natural language processing and text mining features, they tend to lack the ability to concentrate on the targets.
We define this problem of retrieving a subset of documents that are relevant to given targets from large-scale datasets with high recall as search space reduction.
Our solution is to iteratively retrieve the relevant documents utilizing user feedback.
Over multiple iterations, users inspect a topical summary of previously retrieved documents and give feedback, and our system updates targets to better reflect their mental model and retrieve relevant documents through sifting.

In this section, we first formulate the problem of interactive search space reduction, then describe our iterative workflow and algorithm.

\begin{table}[ht!]
	\caption{Key notations used in the paper.}
\vspace{-0.5em}
	\centering
	\begin{tabular}{p{1.15cm}p{6.75cm}}
\toprule
\textbf{Notation} & \textbf{Description}\\
\midrule
$D$    & Given document collection of $n$ documents $\{d_1,\cdots,d_n\}$\\
$W$    & Given keyword dictionary of $m$ keywords $\{w_1,\cdots,w_m\}$\\
$X$    & The $m\times n$ word-document matrix of $D$\\
\midrule
$t$    & Current iteration number\\
$D^{(t)}$      & Set of $n_t$ retrieved documents at the $t$-th iteration, $\{d^{(t)}_1,\cdots,d^{(t)}_{n_t}\}$\\
$s_r^{(t)}(\cdot)$      & Relevance score of document/topic at the $t$-th iteration\\
$s_c^{(t)}(d)$      & Topic score of document $d$ at the $t$-th iteration\\
$G^{(t)}$      & Set of $r_t$ targets at the $t$-th iteration, $\{g^{(t)}_1,\cdots,g^{(t)}_{r_t}\}$\\
$\pmb{G}^{(t)}$      & Set of $r_t$ target vectors for $G^{(t)}$, $\{\pmb{g}^{(t)}_1,\cdots,\pmb{g}^{(t)}_{r_t}\}$\\
$T^{(t)}$ 	   & Set of $k_t$ topics at $t$-th iteration, $\{T^{(t)}_1,\cdots,T^{(t)}_{k_t}\}$\\
$w(T)(w(d))$ & Set of top ten keywords of a topic $T$ (or a document $d$)\\
$d(T)$ & Set of documents that belong to a topic $T$\\
\midrule
$W^{(t)}_+(W^{(t)}_-)$ & Set of good-to-have (bad-to-have) keywords by users at the $t$-th iteration\\
$T^{(t)}_+ (T^{(t)}_-)$ & Set of upvoted (downvoted) topics by users at the $t$-th iteration\\
$D^{(t)}_+(D^{(t)}_-)$ & Set of upvoted (downvoted) documents by users at the $t$-th iteration\\
\midrule
$X^{(t)}$    & The word-document matrix of $D^{(t)}$, $X^{(t)}=X(:,D^{(t)})$ \\
$V^{(t)}$    & The word-topic matrix of $T^{(t)}$\\
$H^{(t)}$    & The topic-document matrix of $T^{(t)}$\\ 
$\pmb{x}_j^{(t)},~\pmb{v}_j^{(t)},~\pmb{h}_j^{(t)}$    & $j$-th column of $X^{(t)}$, $V^{(t)}$, $H^{(t)}$, respectively\\
\midrule
$\mathbb{R^+}$ & The set of nonnegative real numbers \\
$||\cdot ||_F$ & The Frobenius norm\\
$\pmb{e}_i$ & The standard basis vector where $\pmb{e}_i(j)=1$ for $j=i$.\\
$A_{i\cdot}$ & The $i$-th row of matrix $A$\\
$A_{\cdot j}$& The $j$-th column of matrix $A$\\   
$argmax(\pmb{a})$& The index of the largest element in vector $\pmb{a}$\\   
\bottomrule                                        
	\end{tabular}
	\label{tab:notation}%
\vspace{-1em}
\end{table}

\subsection{Problem Formulation and Algorithm Workflow}
\label{sec:workflow}
Given a document collection $D=\{d_1,\cdots,d_n\}$ with $n$ documents, our goal is to retrieve a subset $D^*\subseteq D$ of documents that are relevant with high recall.
Note that we do not limit the number of retrieved documents $|D^*|$, as opposed to traditional information retrieval.
Our iterative approach updates targets $G^{(t)}$ based on user feedbacks and retrieves documents $D^{(t)}$ over iterations $t=1,\cdots,T$.

Our algorithm workflow is outlined in Fig.~\ref{fig:workflow}, with notation listed in Table~\ref{tab:notation}. 
Each iteration consists of three computational steps: target building, sifting, and targeted topic modeling.
An iteration starts with user feedback from its previous iteration.
After exploration of previously retrieved documents and their topics, users can modify keyword queries and/or give positive or negative feedback on topics or documents.
Based on the user input, the \textit{interactive target building} step (Section~\ref{sec:targetmodel}) updates the targets $G^{(t-1)}\rightarrow G^{(t)}$.
Next, the \textit{sifting} step (Section~\ref{sec:sifting}) selects a new set of documents $D^{(t)}$ using the updated targets $G^{(t)}$.
Finally, the \textit{targeted topic modeling} step (Section~\ref{sec:topicmodel}) generates topics $T^{(t)}$ and the system visualizes them.
The users can repeat the iterative process until satisfied.

\subsection{Interactive Target Building Based on User Feedback}
\label{sec:targetmodel}
We represent targets as a set of single keywords (e.g., ``apple'') or keyword compounds (e.g., ``apple, orange'').
The former looks for documents containing the single keyword and the latter looks for those containing all of the keywords in the keyword compound.
In the search space reduction problem, users may not be familiar with their target domains \cite{white2009exploratory}.
Even for domain expert users, constructing a good static query is a challenging task without exploring and understanding given datasets in advance.
Both cases can be solved with interactive target building.
At each iteration, our interactive target building step updates targets based on user feedback.

Different from existing information retrieval approaches that use positive queries, we use negative as well as positive targets.
This allows users to express their complicated mental target model.
For example, the users may be interested in a target, but not interested in a similar concept (e.g., retrieve ``apple, fruit'' and ignore ``orange, fruit'').
Negative targets can also deal with multi-meaning words (e.g., retrieve ``apple, iphone'' and ignore ``apple, fruit'').
In detail, we allow the users to directly update the keyword sets including good-to-have keywords, bad-to-have keywords, stopwords to be ignored.
Stopwords are the words that are not useful in text analysis including too frequent words such as articles, prepositions, and pronouns.
In addition to the commonly used English stopwords, we allow the users to add custom stopwords that are data-specific or domain-specific.
For example, when exploring medical records, ignoring common medical terms may increase the quality of topic modeling and sifting. 
Also, the users can indirectly update the target by giving item-level (documents) or group-level (topics) relevance feedback.

Our approach incorporates seven kinds of user relevance feedback into target building:
\begin{enumerate}[label=\textbf{RF \arabic*}, itemindent=2em]
\setlength{\itemsep}{-.2\baselineskip}
	\item \textit{Edit good-to-have keywords} \label{rf:pw}
	\item \textit{Edit bad-to-have keywords} \label{rf:mw}
	\item \textit{Edit stopwords (words to be ignored)} \label{rf:iw}
	\item \textit{Confirm/upvote topics} \label{rf:pt}
	\item \textit{Reject/downvote topics} \label{rf:mt}
	\item \textit{Confirm/upvote documents} \label{rf:pd}
	\item \textit{Reject/downvote documents} \label{rf:md}
\end{enumerate}
\ref{rf:pw},~\ref{rf:pt},~\ref{rf:pd} are positive relevance feedback indicating that the corresponding words, topics, or documents are relevant to the user's mental target $G$, respectively.
On the contrary, \ref{rf:mw},~\ref{rf:mt},~\ref{rf:md} are negative relevance feedback indicating that the corresponding words, topics, or documents are irrelevant to the user's mental target $G$, respectively.
Lastly, \ref{rf:iw} modifies the set of stopwords, which affects the follow-up topic modeling process described in Section~\ref{sec:topicmodel}.

Given user relevance feedback, we model the targets and their representative vectors as follows:\\
\noindent\fn{TargetModel} computes the targets $G^{(t)}$ and their vectors $\pmb{G}^{(t)}$ at the $t$-th iteration using the user supplied input $(W^{(t)}_+,~W^{(t)}_-,~T^{(t)}_+,~T^{(t)}_-,~D^{(t)}_+,~D^{(t)}_-)$.
The target $G^{(t)}$ consists of positive/negative explicit/implicit parts. 
Users can change the explicit part $G^{(t)}_+,~G^{(t)}_-$ directly through keyword modification.
For implicit part $\bar{G}^{(t)}_+,~\bar{G}^{(t)}_-$, using relevance feedback on a topic or a document, we extract its top keywords and add the keyword compound as an implicit target.

\vspace{-1em}
\begin{algorithm}[H]
	\caption{$[G^{(t)},~\pmb{G}^{(t)}] = \fn{TargetModel}(W^{(t)}_+,~W^{(t)}_-,~T^{(t)}_+,~T^{(t)}_-,~D^{(t)}_+,~D^{(t)}_-)$}
	\begin{algorithmic} 
	\STATE $G^{(t)}_+ = W^{(t)}_+$; $\pmb{G}^{(t)}_+ = \{\frac{\pmb{a}}{\|~\pmb{a}\|_2}:\pmb{a}=\sum_{w\in{g}}{\pmb{e}_{w}},~g\in G^{(t)}_+\}$
	\STATE $G^{(t)}_- = W^{(t)}_-$; $\pmb{G}^{(t)}_- = \{\frac{\pmb{a}}{\|~\pmb{a}\|_2}:\pmb{a}=\sum_{w\in{g}}{\pmb{e}_{w}},~g\in G^{(t)}_-\}$
	\STATE $\bar{G}^{(t)}_+ =\{w(T_j^{(t-1)}):T_j^{(t-1)}\in T^{(t)}_+\} \cup \{w(d_j^{(t-1)}):d_j^{(t-1)}\in D^{(t)}_+\}$
	\STATE $\bar{\pmb{G}}^{(t)}_+ = \{\pmb{v}_j^{(t-1)}:T_j^{(t-1)}\in T^{(t)}_+\} \cup \{\pmb{x}_j^{(t-1)}:d_j^{(t-1)}\in D^{(t)}_+\}$
	\STATE $\bar{G}^{(t)}_- =\{w(T_j^{(t-1)}):T_j^{(t-1)}\in T^{(t)}_-\} \cup \{w(d_j^{(t-1)}):d_j^{(t-1)}\in D^{(t)}_-\}$
	\STATE $\bar{\pmb{G}}^{(t)}_- = \{\pmb{v}_j^{(t-1)}:T_j^{(t-1)}\in T^{(t)}_-\} \cup \{\pmb{x}_j^{(t-1)}:d_j^{(t-1)}\in D^{(t)}_-\}$
	\STATE $G^{(t)} = (G^{(t)}_+,~G^{(t)}_-,~\bar{G}^{(t)}_+,~\bar{G}^{(t)}_-)$; $\pmb{G}^{(t)} = (\pmb{G}^{(t)}_+,~\pmb{G}^{(t)}_-,~\bar{\pmb{G}}^{(t)}_+,~\bar{\pmb{G}}^{(t)}_-)$
	\end{algorithmic}
\end{algorithm}

\vspace{-1em}
\subsubsection{Keyword Suggestion}
\label{sec:keyword_recomm}
Manually entering keywords can be burdensome.
To this end, we recommend candidates for the good-to-have and bad-to-have keyword sets in real time.
Candidate recommendation is based on similarities with the current good-to-have and bad-to-have keyword sets.
Similarities between words can be calculated by  several distance measures.
Among them, we adopt the vector-space model of word representation~\cite{mikolov2013efficient}.
To learn word vectors, we use empirical pointwise mutual information (ePMI), which measures co-occurrence between word pairs.
The ePMI score between the word pair $(w_i,w_j)$ is defined as:
\vspace{-1em}
\begin{equation}
ePMI(w_i,w_j)=\log\Big(\frac{\#(w_i,w_j)\cdot N}{\#(w_i)\cdot\#(w_j)}\Big),
\label{eqn:word_vec}
\vspace{-1em}
\end{equation}

\noindent where $N$ denotes the total number of the word co-occurring word pairs; and $\#(w_i,w_j)$ and $\#(w_i)$ denote the number of occurrences of the word pair $(w_i,w_j)$ and the single word $w_i$, respectively.
As suggested by ~\cite{levy2014neural}, we first construct a matrix $P\in \mathbb{R}^{m\times m}$ where $P_{i,j} = ePMI(w_i, w_j)$, perform low-rank matrix factorization on $P$, and use the left factor as the vector representations of words after $l2$-normalization.
We computed word vectors for each dataset to obtain dataset-specific word similarities, but pre-trained word vectors using word2vec~\cite{mikolov2013efficient} or Glove~\cite{pennington2014glove} can be used in our algorithm.

\subsection{Sifting Documents and Words}
\label{sec:sifting}
After the target modeling step, we retrieve a new set of documents using the updated targets.
We provide two retrieval options: hard filtering by target keywords and soft sifting.

\fn{HardSift} throws out documents that contain one of the negative target elements or their nearest words and retrieves documents that contain one of the target elements or their nearest words.
One of nearest words of a word $w$ is denoted by $sim(w)$.
Note that we apply negative feedback first and positive feedback later to take a conservative approach in filtering out documents.

\vspace{-1em}
\begin{algorithm}[H]
	\caption{$D^{(t)} = \fn{HardSift}(G^{(t)},~T^{(t)}_{+},~T^{(t)}_{-},~D^{(t)}_{+},~D^{(t)}_{-},~D^{(t-1)})$}
	\begin{algorithmic} 
	\STATE $D^{(t)} = D^{(t-1)} \bigcup_{j=1}^{|G^{(t)}_+|}\{d_i\in D : \forall w \in g_j (\in G^{(t)}_+),~d_i~\text{has}~sim(w) \}$
	\STATE $D^{(t)} = D^{(t)}\big\backslash_{j=1}^{|G^{(t)}_-|}\{d_i\in D^{(t)} : \forall w \in g_j (\in G^{(t)}_-),\thinspace d_i~ \text{has}~sim(w)\}$
    \STATE $D^{(t)} = D^{(t)} \big\backslash_{T_j^{(t-1)} \in T^{(t)}_{-}}d(T_j^{(t-1)})$
    \STATE $D^{(t)} = D^{(t)} \big\backslash_{d_j^{(t-1)} \in D^{(t)}_{-}}\{d_i\in D^{(t-1)} : (\pmb{x}_i \cdot \pmb{x}_j^{(t-1)}) > \delta\}$
    \STATE $D^{(t)} = D^{(t)} \bigcup_{T_j^{(t-1)} \in T^{(t)}_{+}}\{d_i\in D : (\pmb{x}_i \cdot \pmb{v}_j^{(t-1)}) > \delta\}$
    \STATE $D^{(t)} = D^{(t)} \bigcup_{d_j^{(t-1)} \in D^{(t)}_{+}}\{d_i\in D : (\pmb{x}_i \cdot \pmb{x}_j^{(t-1)}) > \delta\}$
	\end{algorithmic}
\end{algorithm}
\vspace{-1em}

\fn{SoftSift} incorporates a relevance score model to rank documents by how similar they are to the explicit and implicit targets.
The relevance score of a document with respect to a target $g$ is calculated as cosine similarity between its target vector $\pmb{g}$ and the document vector $\pmb{x}$, $(\pmb{x} \cdot \pmb{g})$.
All target vectors and document vectors are l2-normalized.
To calculate the final relevance score of a document, we take a weighted average of its previous relevance score and its relevance scores with respect to positive and negative feedbacks at the current iteration.
To put more emphasis on recall than precision, we use smaller weight for negative feedback score than positive feedback score, i.e. $\beta>\gamma$.

\vspace{-1em}
\begin{algorithm}[H]
	\caption{$D^{(t)} = \fn{SoftSift}(\pmb{G}^{(t)},~D)$}
	\begin{algorithmic} 
	\STATE $\alpha,~\beta,~\gamma$ is parameters for balancing previous scores, positive feedback, and negative feedback, respectively.
	\STATE $\delta$ is the threshold for the soft mode.
	\FOR{$d_i \in D$}
		\STATE $s_{r+}^{(t)}(d_i) = mean_{\pmb{g}_+^{(t)} \in \pmb{G}^{(t)}_+\cup\bar{\pmb{G}}^{(t)}_+} (\pmb{x}_i \cdot \pmb{g}_+^{(t)})$
		\STATE $s_{r-}^{(t)}(d_i) = mean_{\pmb{g}_-^{(t)} \in \pmb{G}^{(t)}_-\cup\bar{\pmb{G}}^{(t)}_-} (\pmb{x}_i \cdot \pmb{g}_-^{(t)})$
		\STATE $s_r^{(t)}(d_i) = \alpha s_r^{(t-1)}(d_i) + \beta s_{r+}^{(t)}(d_i)  - \gamma s_{r-}^{(t)}(d_i)$
	\ENDFOR
	\STATE $D^{(t)} = \{d_i\in D:s_r^{(t)}(d_i)> \delta\}$
	\end{algorithmic}
\end{algorithm}

\vspace{-1em}
\subsection{Targeted Topic Modeling}
\label{sec:topicmodel}
The last step of an iteration is targeted topic modeling.
Targeted topic modeling finds a target-specific topical summary of documents that are retrieved from the previous sifting step.
The calculated topics and their representative documents are visualized to the users so that they can easily understand what kind of documents are retrieved at the current iteration and perform relevance feedback for the next iteration.

In this section, we explain nonnegative matrix factorization (NMF)~\cite{jogo2015} in the topic modeling context ~\cite{Du2017,Drake2017} and our targeted topic modeling algorithm based on NMF with additional constraints.

\subsubsection{Background: NMF for Topic Modeling}
Given a nonnegative matrix $X\in \mathbb{R}_+^{m\times n}$, NMF approximates X as a product of nonnegative factor matrices $V\in \mathbb{R}_+^{m\times k}$ and $H\in \mathbb{R}_+^{k\times n}$, i.e., $X\approx VH$, with $k\ll\min(m,n)$.
This can be solved by optimizing the following formula:

\vspace{-1.2em}
\begin{equation}
\min_{\{V,H\}\geq 0}||X-VH||^2_F.
\label{eqn:nmf}
\vspace{-0.7em}
\end{equation}

\noindent In the topic modeling context, $X$ is a word-document matrix where $X_{\cdot j}$ (the $j$-th column vector of $X$) is a bag-of-words representation of $j$-th document over $m$ keywords.
$X$ is based on TF-IDF representation of the document set and usually normalized with $l2$-norm.
$k$ is set to be the number of topics.
Factor matrices $V$ and $H$ represent word-topic and topic-document relationships, respectively.
$V_{\cdot i}$ represents the $i$-th topic as a distribution over words.
Large values in $V_{\cdot i}$ indicate that the corresponding keywords are strongly associated with the $i$-th topic.
$H_{\cdot j}$ represents the $j$-th document $d_j$ as a weighted combination of topics.
The $j$-th document $d_j$ belongs to the $i$-th topic if the $i$-th element of $H_{\cdot j}$ is its maximum, i.e., $argmax(H_{\cdot j}=i)$.
We denote the $i$-th topic as $T_i$ and define it by its word distribution vector ($w(T_i)=V_{\cdot i}$) and the documents that belong to it ($d(T_i)=\{d_j|argmax(H_{\cdot j}\}$).

\subsubsection{Targeted Topic Modeling using NMF}
To reflect a target built by users into the topic modeling process, we introduce an additional constraint term to the standard NMF formula, Eqn.~\ref{eqn:nmf}, as follows:

\vspace{-1em}
\begin{equation}
\min_{\{V,H\}\geq 0}||X-VH||^2_F+\rho||M\circ V-V_G||^2_F,
\label{eqn:steer}
\vspace{-0.7em}
\end{equation}

\noindent where $\circ$ is an elementwise multiplication.
The additional term forces certain topics' word representation $V$ to be similar to the corresponding target elements $V_G$ with the help of masking coefficient matrix $M$.
The parameter $\rho$ controls the balance between the original term and the additional term.
Bigger $\rho$ results in stronger incorporation of the target in topic modeling.
That is, the bigger the rho is, the closer the topics become to the targets at the expense of becoming less truthful representation of data.
When $\rho=0$, it is equivalent to the standard topic modeling.
Also, $\rho$ is inversely proportional to the number of positive targets.
To compute $M$ and $V_G$, for each positive target vector $\pmb{g}_i \in \pmb{G}_+ \cup \bar{\pmb{G}}_+$, find its closest topic vector, which we define as $\pmb{v}_i^*=argmax_{\pmb{v}_j}(\pmb{v}_j \cdot \pmb{g}_i)$.
We set $(V_G)_{j\cdot}=mean_{\pmb{v}_i^*=\pmb{v}_j}(\pmb{g}_i)$ and 
$M_{j\cdot}=1$ if $|\{\pmb{g}_i:\pmb{v}_i^*=\pmb{v}_j\}|>0$.

The detailed algorithm at the $t$-th iteration is as follows:\\
\noindent\fn{TargetedTopicModel} applies a constrained NMF algorithm on the current document set $D^{(t)}$ and the current targets $G^{(t)}$ to compute $k_t$ number of topics $T^{(t)}$.
Additionally, we calculate each topic's relevance score with respect to the targets. Note that $rank(w,T_i)$ calculates the rank of a word $w$ within the topic $T_i$'s topic vector $\pmb{v}_i$.
For speedup, we use a fast rank-2 NMF~\cite{kdd2013} algorithm to initialize $V$ and $H$ in Eqn.~\ref{eqn:nmf}.

\vspace{-1em}
\begin{algorithm}[H]
	\caption{$[T^{(t)},~s_r^{(t)},~s_c^{(t)}] = \fn{TargetedTopicModel}(X^{(t)},~k_t,~G^{(t)})$}
	\begin{algorithmic} 
	\STATE Generate $V^{(t)}_G$, $M^{(t)}$ and solve
	\STATE $\min_{\{V^{(t)},H^{(t)}\}\geq 0}||X^{(t)}-V^{(t)}H^{(t)}||^2_F+\rho||M^{(t)}\circ V^{(t)}-V^{(t)}_G||^2_F$
	\STATE $s_r^{(t)}(T^{(t)}) = 1 - min_{g\in G_{+}^{(t)}} \big( mean_{w\in g} \dfrac{rank(w,T^{(t)})}{|W|} \big)$
	\STATE $s_c^{(t)}(d_j^{(t)}) = max(\pmb{h}_j) / sum(\pmb{h}_j)$
	\end{algorithmic}
\end{algorithm}

\section{System}
\label{sec:system}
In this section, we present TopicSifter, our interactive document search space reduction system. 
Our visualization system is tightly integrated with the underling algorithms described in Section~\ref{sec:technique} to support various user feedback interactions listed in Section~\ref{sec:targetmodel}.

TopicSifter is designed to meet these design goals:

\begin{enumerate}[label=\textbf{\arabic*.}, itemindent=0em]
\vspace{-1em}
	\setlength{\itemsep}{-.2\baselineskip}
	\item \textbf{Given targets, retrieve relevant documents with high recall:} TopicSifter should retrieve documents that are relevant to targets. \label{dg:retrieve}
	\item \textbf{Show summary and details of sifted documents:} TopicSifter should provide a topical summary and details of retrieved documents to help users understand them. \label{dg:topicdetail}
	\item \textbf{Support positive and negative feedback:} Users should be able to positive and negative relevance of both keywords, documents, and topics (Supporting \ref{rf:pw}-\ref{rf:md}). \label{dg:doctopic}
	\item \textbf{Modify targets over iterations:} TopicSifter should allow users to update targets easily and iteratively. \label{dg:target}
	\item \textbf{Observe changes between iterations:} TopicSifter should show differences in retrieved documents between iterations. \label{dg:change}
	\item \textbf{Export results for further analysis:} TopicSifter is designed for one step of a complex text analysis workflow. Users should be able to export the retrieved documents for in-depth analyses. \label{dg:export}
\vspace{-1.5em}
\end{enumerate}

TopicSifter consists of a web-based visualization interface using D3.js and a backend system in Python and MATLAB using the Django framework.

\begin{figure}[t]
	\centering
	\includegraphics[width=\linewidth]{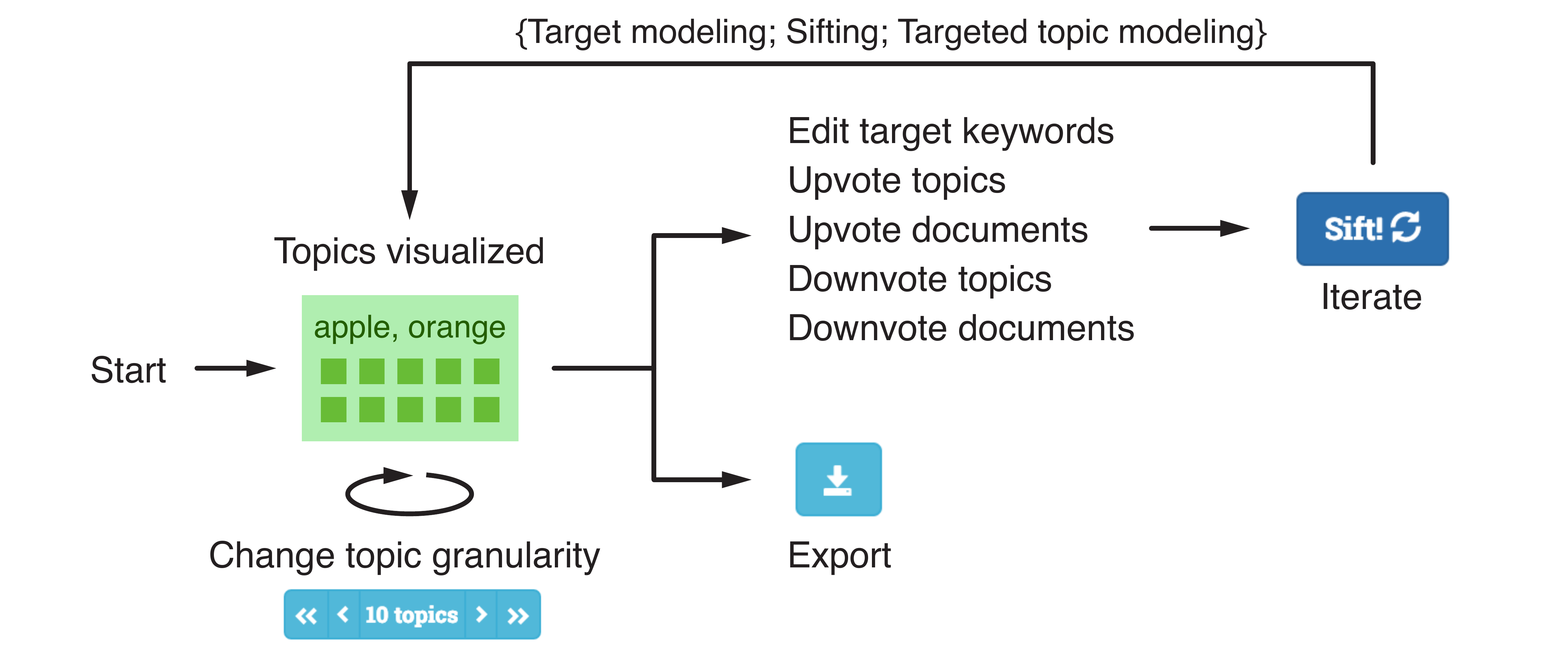}
	\caption{TopicSifter workflow. Users can provide feedback before clicking the blue button to move to the next iteration. Results can be saved by clicking the export button.}
	\label{fig:system_workflow}
\vspace{-1.5em}
\end{figure}

\subsection{System Overview}
TopicSifter consists of three panels: (1) the control panel, (2) the main view, and (3) the detail panel (Fig.~\ref{fig:teaser}).
The control panel contains the keyword module to modify good-to-have words, bad-to-have words, and stopwords (supporting \ref{rf:pw}, \ref{rf:mw}, \ref{rf:iw}) and control buttons to update the main view.
The main view shows the sifting status and the topical overview of retrieved documents at the current iteration and allows the users to upvote or downvote topics and documents (supporting \ref{rf:pt},~\ref{rf:mt},~\ref{rf:pd},~\ref{rf:md}). 
The relevance feedback on words, topics, and documents will be reflected on the next iteration (Fig.~\ref{fig:system_workflow}).
Lastly, the detail panel has the document table to show additional detail of all documents and the history view to show historical trends over iterations.
The width of each panel is adjustable by dragging the divider in order to allocate more or less space to the panel.
The system design is shown in Fig.~\ref{fig:teaser}.

The users follow the workflow in Fig.~\ref{fig:system_workflow}.
Each iteration starts with the users exploring the retrieved documents and their topics in the main view.
To give relevance feedback, the users can modify keyword sets in the control panel or upvote/downvote topics and documents in the main view.
They can export the results or move on to the next iteration using buttons in the control panel.

\subsection{Control Panel}
The users can utilize the control panel to update the main view.
The control panel contains the keyword input module and the control buttons.
The keyword input module shows current set of good-to-have keywords $W^{(t)}_+$, bad-to-have keywords $W^{(t)}_-$, and stopwords and allows users to modify them (\ref{rf:pw}, \ref{rf:mw}, \ref{rf:iw}). \\

\vspace{-0.5mm}
\noindent\textbf{\textsf{Keyword Input Module}}

\begin{figure}[t]
	\centering
	\includegraphics[width=\linewidth]{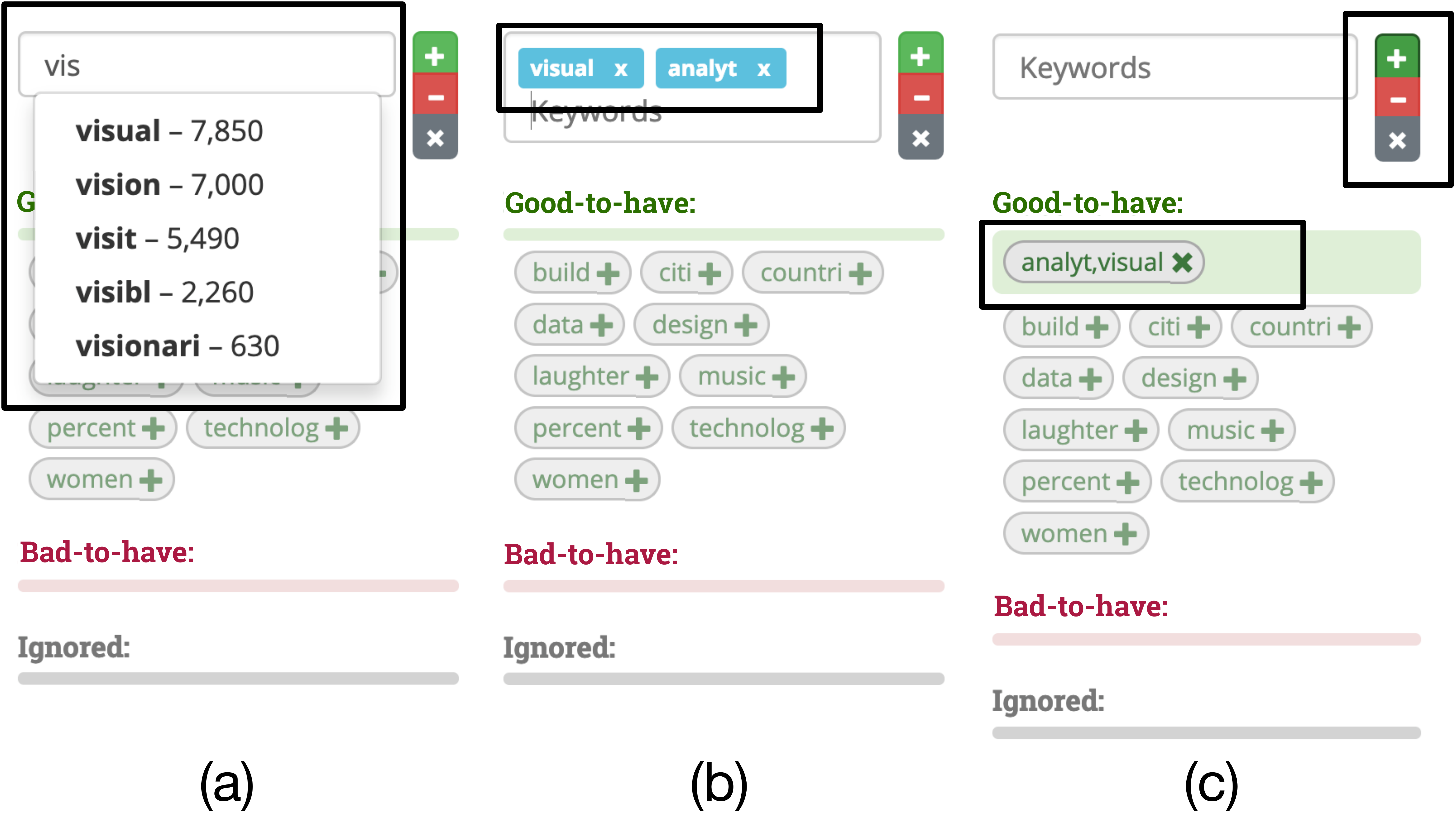}
\vspace{-1.5em}
	\caption{Users can add good-to-have words, bad-to-have words, and stopwords in the control panel. 
(a) While typing, partially matched keywords are ranked by frequency and shown in a pop-up list. 
(b) Multi-word compound is supported. 
(c) Clicking the green button adds the entered keyword compound into the good-to-have set.}
	\label{fig:query}
\vspace{-1.5em}
\end{figure}

\noindent In the keyword input module (Fig.~\ref{fig:teaser}(d)), the users can add new keywords using an input text box or see current keyword lists for good-to-have keywords, bad-to-have keywords, and stopwords.
To add a keyword, users can enter the keyword in the input text box.
While typing, possible matching keywords in the dictionary $W$ is listed in the pop-up as shown in Fig.~\ref{fig:query}(a).
The list is sorted by word frequency and updated as the user types more letters.
After selecting one of the keywords in the pop-up list, the users can either enter the keyword as a single keyword (e.g., ``visual'') or build a keyword compound (e.g., ``visual'' AND ``analyt(ic)'' in Fig.~\ref{fig:query}(b)).
By clicking one of the green, red, or gray buttons in Fig.~\ref{fig:query}(c), the entered keyword or keyword compound is added in the good-to-have keyword list $W^{(t)}_+$, the bad-to-have keyword list $W^{(t)}_-$, or the stopword list, respectively.
Keywords or keyword compounds in the keyword list is visualized as word buttons inside the colored areas (good-to-have: green, bad-to-have: red, stopword: gray) as in Fig.~\ref{fig:query}(c).
In order to remove a keyword or a keyword compound, the users can click the $\times$ icon on the keyword button.

As discussed in Section~\ref{sec:keyword_recomm}, our technique suggests additional keywords based on the current set of keywords. 
The keywords recommended for good-to-have or bad-to-have lists are visualized under the corresponding keyword list as keyword buttons with dashed borders with a + icon.
The users can add one of suggested keywords by clicking the + icon.
The recommended keywords are updated in real time as the users add or remove keywords to the keyword lists.

\vspace{1mm}
\noindent\textbf{\textsf{Changing the number of topics}}
\includegraphics[height=0.17in]{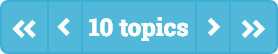}

\noindent Users can change the topic granularity by increasing or decreasing the number of topics using the button group in the control panel.
When the generated topics are too fine-grained or too coarse-grained, giving relevance feedback can be problematic.
For example, the user wants to give feedback on all ``fruit'' related topics, but there are too many fine-grained ``fruit'' related topics to interact with.
On the other hand, the user may be interested in part of a topic (e.g., like ``apple, mac'' part from ``apple, mac, fruit'', but not ``apple, fruit'' part). 
The number of current topics is shown in the middle part of the button group.
The users can click the buttons to decrease the number of topics by -5($\ll$), -1($<$), or increase it by +1($>$), +5($\gg$).
Note that a new set of topics is generated using the same retrieved document subset.
The visual update after changing the number of topics is fast since this happens within an iteration without triggering the target building, sifting, and targeted topic modeling processes (Fig.~\ref{fig:system_workflow}).

\vspace{1mm}
\noindent\textbf{\textsf{Sift Button}}
\includegraphics[height=0.17in]{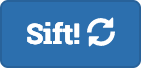}

\noindent The users can run our backend algorithms by clicking the sift button. 
After modifying good-to-have keywords, bad-to-have keywords, and stopwords (\ref{rf:pw}, \ref{rf:mw}, \ref{rf:iw}) and upvoting or downvoting topics and documents (\ref{rf:pt}, \ref{rf:mt}, \ref{rf:pd}, \ref{rf:md}), the users move on to the next iteration.
The sift button triggers the target building, sifting, and targeted topic modeling processes to retrieve a new set of documents and visualize their topic summary.
This process is shown in Fig.~\ref{fig:system_workflow}.

\vspace{1mm}
\noindent\textbf{\textsf{Export Button}}
\includegraphics[height=0.17in]{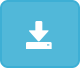}

\noindent The users can export the results using the export button.
When the users are satisfied with the retrieved documents after multiple iterations, our system provides an  option to save the results.
The results are saved as a JSON file including targets, topics, and IDs, topic membership, and relevance scores of retrieved documents.

\subsection{Main View}
The main view will visualize topic summary of retrieved documents at the current iteration along with the sifting status bar to show the difference between the current iteration and the previous iteration.
In the topic visualization, the users can upvote or downvote topics and documents to indicate that they are relevant to targets or not (\ref{rf:pt}, \ref{rf:mt}, \ref{rf:pd}, \ref{rf:md}).\\

\begin{figure}[t]
	\centering
	\includegraphics[width=\linewidth]{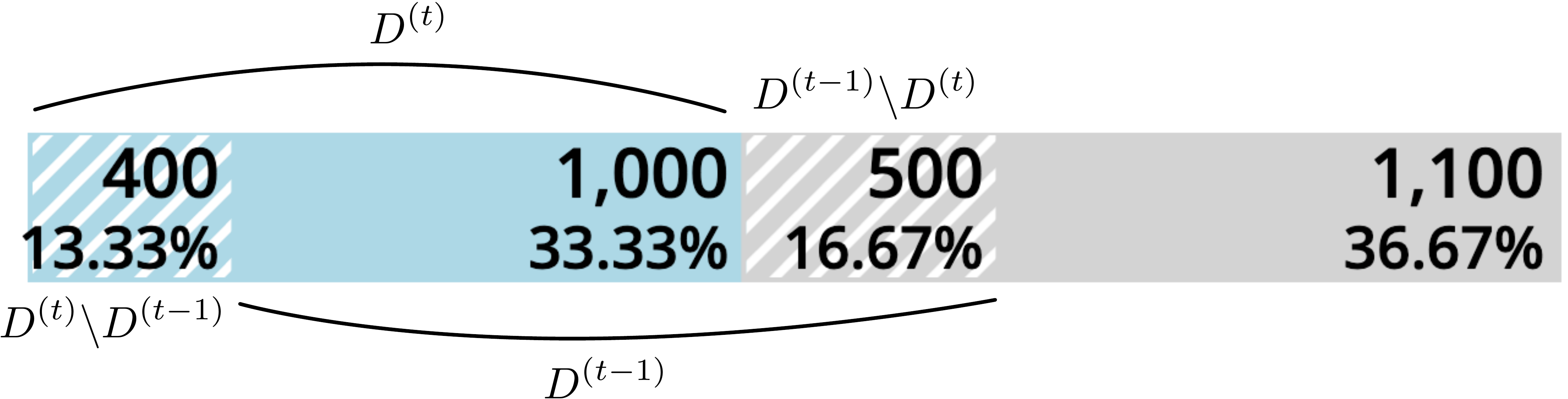}
\vspace{-1.7em}
	\caption{The status bar chart that shows sifting status at the current $t$-th iteration. 
Blue bars represent retrieved documents $D^{(t)}$ at the $t$-th iteration. 
Gray bars represent the rest (sifted out) documents in the corpus, i.e., $D\backslash D^{(t)}$. 
Patterned bars indicate changes from the $(t-1)$-th iteration.}
	\label{fig:bar}
\vspace{-1.5em}
\end{figure}

\noindent\textbf{\textsf{Status Bar}}

\noindent Fig.~\ref{fig:bar} shows the status bar chart.
The total length of all bars represents the total number of documents in the dataset.
The total length of blue bars represents the number of retrieved documents at the current iteration, $t$, while the total length of gray bars represents the number of sifted out documents.
Solid-colored bars represent documents that stay retrieved (solid blue) or stay sifted out (solid gray) between the previous $(t-1)$-th iteration and the current $t$-th iteration.
Patterned bars represent document status changes from the previous iteration, $t-1$.
In detail, the blue patterned bar represents incoming documents that were not retrieved at the $(t-1)$-th iteration but retrieved at the $t$-th iteration.
The gray patterned bar represents outgoing documents that were retrieved at the $(t-1)$-th iteration but sifted out at the $t$-th iteration.
Longer patterned bars indicate interactions at the $t$-th iteration have resulted in a larger change in retrieved documents.\\

\begin{figure}[t]
	\centering
	\includegraphics[width=\linewidth]{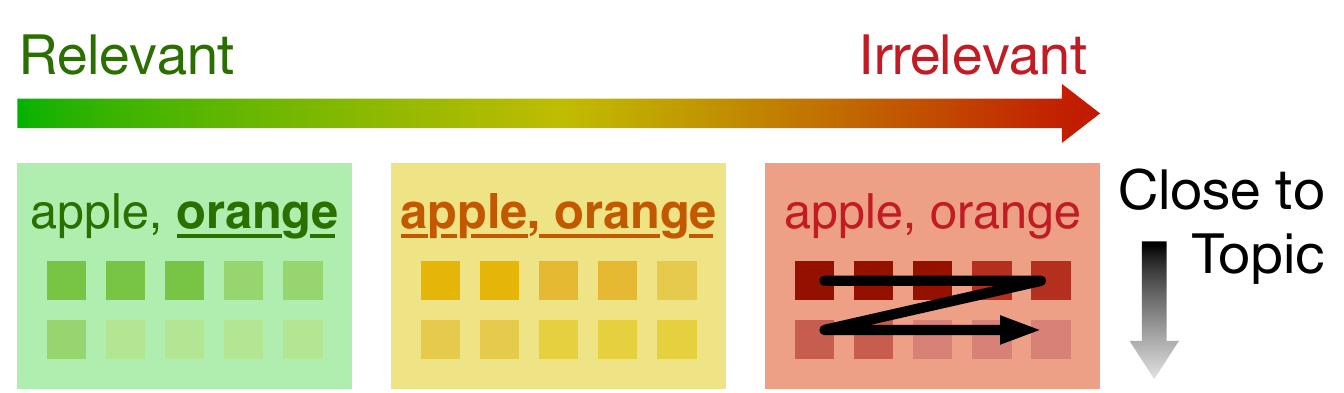}
\vspace{-1.5em}
	\caption{Visual encoding of topic cells and their representative documents. 
Target-relevancy of topics are encoded by their color hue (green to red).
Topic-closeness of documents are encoded by their color lightness (dark to light) and positions (top-left to bottom-right). 
Topic change from the previous iteration is indicated by new keywords highlighted as bold and underlined.}
	\label{fig:topic_color}
\vspace{-1.1em}
\end{figure}

\begin{figure}[t]
	\centering
	\includegraphics[width=\linewidth]{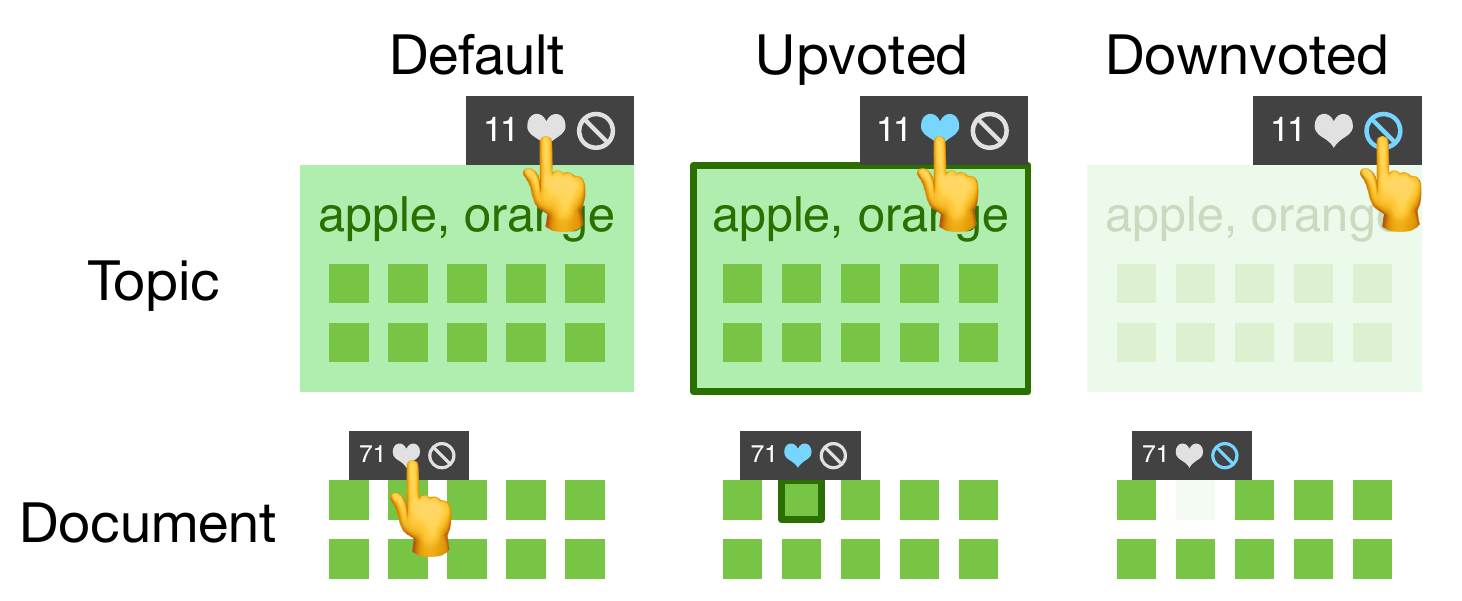}
\vspace{-1.8em}
	\caption{User interaction for positive and negative feedback. 
Users can click the upvote (or downvote) button in the pop-up menu of a topic or a document to indicate (ir-)relevancy. 
Highlight with border means upvoted and white out means downvoted.}
	\label{fig:feedback_ui}
\vspace{-2.5em}
\end{figure}

\noindent\textbf{\textsf{Topic Visualization}}

\noindent Topics computed from the retrieved documents are visualized as rectangular cells (Fig.~\ref{fig:teaser}).
On top of each topic cell, its top ten keywords are shown, and its representative documents are visualized as small squares.
The sizes of cells are proportional to the number of retrieved documents that belong to each topic.
The layout of cells is calculated by D3's built-in treemap algorithm.
The color hues of topic cells represent how relevant each topic is to the target ($s_r^{(t)}(T)$) from green (relevant) to red (irrelevant) as in Fig.~\ref{fig:topic_color}.
The color hue of each topic is shared by its keywords and its documents.
If a topic has changed from the previous iteration, new representative keywords are highlighted as bold and underlined (The yellow topic in Fig.~\ref{fig:topic_color}).
For topic cells with narrow width, the users can hover over top keywords to see the full list of keywords.
To give positive (or negative) feedback to topics, the users can click the menu button on the top right corner of each topic cells to open a pop-up menu with upvote and downvote button (Fig.~\ref{fig:feedback_ui}).

The number of representative documents that are visualized as squares in a topic cell are determined by the size of the cell.
Our system picks documents to be visualized by how close the documents are to its topic ($s_c^{(t)}(d)$) since they are more representative of the topic.
The color lightness of document squares represents how close each document is to its topic from dark (close) to light (less close).
The positions of document squares are also sorted by closeness to their topics from top-left to bottom-right (Fig.~\ref{fig:topic_color}).
To see the detail of a document, the users can hover over the square to see its document ID in a pop-up menu or click the square to see its detail in the document table in the detail panel.
Users are able to give positive (or negative) feedback to documents to indicate that they are relevant (or irrelevant) to their mental targets by toggling the upvote (or downvote) button in the pop-up menu of each document square as in Fig.~\ref{fig:feedback_ui}.
Upvoted topics and documents are highlighted with border and downvoted topics and documents are whited out.

\begin{figure}[t]
	\centering
	\includegraphics[width=0.8\linewidth]{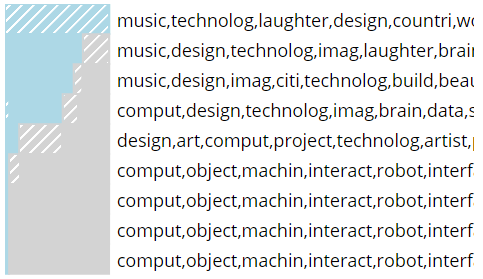}
\vspace{-0.7em}
	\caption{The history view. 
Stacked bars on the left show sifting status from top (old) to bottom (new). 
In each row, blue bars represent retrieved documents at an iteration and gray bars represent the rest (sifted out) documents in the corpus at the same iteration. 
Patterned bars indicate changes from the previous iteration. 
Keyword summary on the right shows the topical progression of retrieved documents over iterations.}
	\label{fig:history}
\vspace{-1.5em}
\end{figure}

\begin{figure*}[t]
	\centering
	\includegraphics[width=\linewidth]{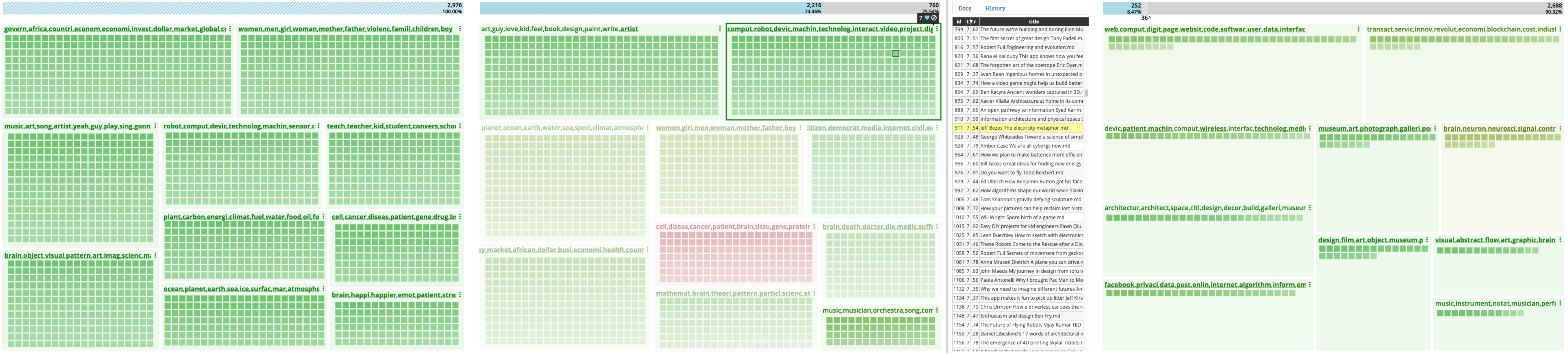}
\vspace{-1.8em}
	\caption{Exploring the TED dataset. 
The initial iteration (left) shows topic summary of all documents. 
After adding ``art, technology'' to the good-to-have list, the user upvotes an interesting document and downvotes another (middle). 
Documents are further sifted (right).}
	\label{fig:ted-s1}
\vspace{-1em}
\end{figure*}

\subsection{Detail Panel}
The detail panel has two tabs to toggle between the document table view and the history view.
The document table view shows the list of all documents $D$ and their raw text details.
The history view shows the history of previous iterations to keep track of the iterative sifting process. \\

\noindent\textbf{\textsf{Document Table}}

\noindent The document table shows additional information of all documents in the dataset, i.e., $D$.
Each row of the table shows document details such as document IDs, titles, raw texts, etc, along with their topic memberships and topic-relevance scores.
The document table is linked with the topic visualization.
Hovering over a document square highlights the corresponding table row, and vice versa. 
Column fields may vary depending on datasets used.
The raw texts can be long, so our system does not show them by default, but a row can be expanded to show the raw text when clicked.
One challenge is that rendering all document rows are impractical in our large-scale text analytics setting.
To solve this, we use Clusterize.js\footnote{Available at: \url{https://clusterize.js.org/}} library to render currently visible rows only and reuse those HTML elements when the table is scrolled.
Another challenge is navigating and scrolling through tens of thousands of rows.
For easy navigation, when a document square in the main view is right-clicked, the document table automatically scrolls to the corresponding row.\\

\noindent\textbf{\textsf{History View}}

\noindent Fig.~\ref{fig:history} shows the history view, which contains a stacked bar chart (left) and the keyword summary history (right).
The stacked bar chart shows all the visualized status bars from previous iterations.
It can reveal changes per iteration and if the sifting results became stable.
The keyword summary history shows top keywords for retrieved documents at each previous iteration.
Users can observe whether their interactions have resulted as expected.

\section{Evaluation}
\label{sec:usecase}
\begin{figure}[t]
	\centering
\vspace{-1em}
	\includegraphics[width=\linewidth]{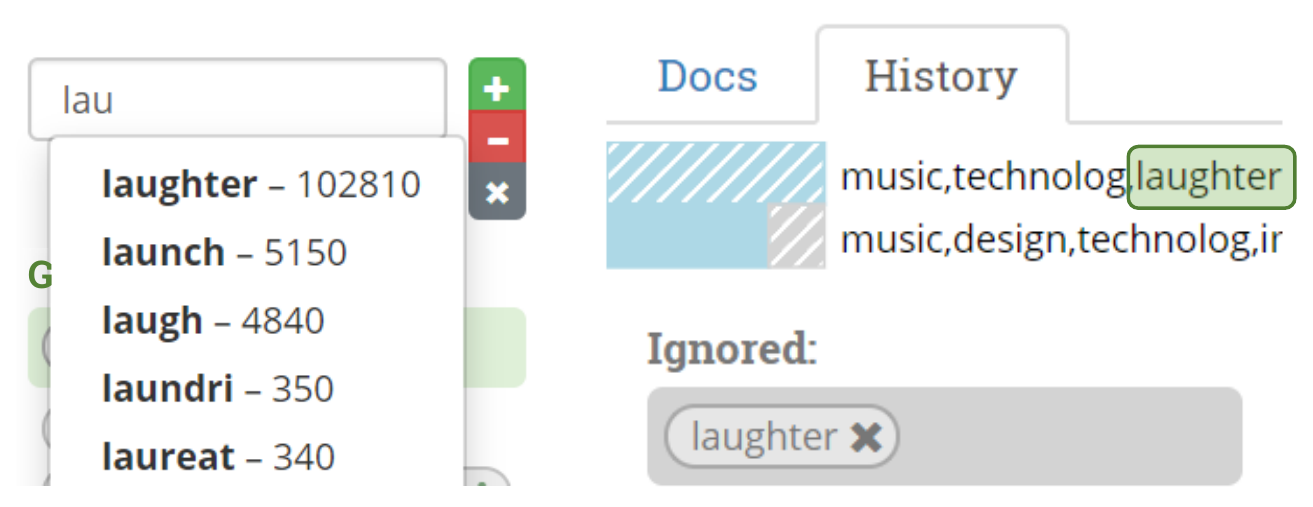}
\vspace{-1.9em}
	\caption{A stopword is detected from the history view.}
	\label{fig:ted-s24}
\vspace{-1.5em}
\end{figure}

In this section, we provide quantitative evaluation utilizing simulated user feedback with a labeled dataset.
Also, we show use cases to illustrate the usefulness of TopicSifter for search space reduction using two datasets: a TED dataset and Twitter dataset.

\subsection{Dataset Description}
The 20 Newsgroup dataset\footnote{Source: \url{http://qwone.com/~jason/20Newsgroups}} is a collection of 19.8K newsgroup documents partitioned into 20 categories.
The size of dictionary is 128K.
The TED talk transcript dataset\footnote{Source: \url{https://github.com/saranyan/TED-Talks}} contains 2,896 documents that are transcribed from the English TED talk videos.
The talks are about various topics including technology, education, etc.
The size of dictionary is 18,275.
The contents of the documents are spoken languages in a subtitle-like style.
The twitter dataset\footnote{Source: \url{https://archive.org/details/twitter\_cikm\_2010}} was originally explored by \cite{cheng2010you}. 
We use part of the data containing 500K tweets.
After removing the documents with less than five words, we are left with 300K documents and 32.3K words.
We applied the Porter stemming algorithm \cite{porter1980algorithm} for pre-processing and built the TF-IDF matrices for the datasets.

\subsection{Quantitative Evaluation}
In this section, we present results of a study simulating user input to test the effectiveness of our technique.

\subsubsection{Experiment Setup}
To simulate user relevance feedback, we used the 20 Newsgroup dataset which has category labels.
Among 20 categories, we chose two labels ``rec.sport.baseball'' (989 documents) and ``rec.sport.hockey'' (993 documents) as relevant/true labels, which is about 10\% of the entire dataset.

First, we entered ``game, team, player, play'', which were four most representative keywords from documents from the two categories, as initial target words.
At each iteration, we select two documents or topics to give relevance feedback on (upvote or downvote based on the true label).
We compared six strategies: 1) upvote two true documents ($+_d$), 2) upvote two true topics ($+_T$), 3) downvote two false documents ($-_d$), 4) downvote two false topics ($-_T$), 5) upvote a true document and downvote a false document ($\pm_d$), 6) upvote a true topic and downvote a false topic ($\pm_T$).

\subsubsection{Results}
Table~\ref{tab:exp} summarizes the performance of different feedback strategies at the 10-th iteration, averaged over three runs.
We used four measures: precision, recall, F1-score (the harmonic mean of precision and recall), and PRES~\cite{magdy2010pres}, which is a recall-oriented measure.
For each strategy, we tried parameters from $\alpha \in \{0.4, 0.5, 0.6, 0.7\}$, $\beta \in \{0.4, 0.5, 0.6\}$, $\gamma \in \{0, 0.1, 0.2\}$ where $\alpha+\beta-\gamma=1$ and chose the combination with best F1 score.
For the sifting threshold, we used $\delta=0.04$.
All strategies converged after 4-6 iterations.

Simulating positive feedback showed higher recall and lower precision than negative feedback.
Performing both positive and negative feedback showed better or comparable scores than performing only positive feedbacks, which advocates our novel negative targeting.
In addition, positive topic-level feedbacks ($+_T,~\pm_T$) outperformed the others in F1 and PRES scores.
This validates that our topic-level relevance feedback is beneficial in search space reduction.

\begin{table}[]
	\caption{Retrieval performance of relevance feedback strategies with their parameter settings. Scores are averaged over three runs. Best scores are highlighted.}
\vspace{-0.5em}
	\centering
	\begin{tabular}{@{}c|cccccc@{}}
\toprule
                 & $+_d$ & $+_T$ & $-_d$ & $-_T$ & $\pm_d$ & $\pm_T$ \\
\midrule
Precision        & 0.798   & 0.670     & 0.825     & \textbf{0.880}       & 0.794       & 0.808         \\
Recall           & 0.667   & \textbf{0.827}     & 0.596     & 0.415       & 0.669       & 0.754         \\
F1          & 0.727   & 0.740     & 0.692     & 0.564       & 0.726       & \textbf{0.780}         \\
PRES~\cite{magdy2010pres}     & -0.095  & 0.680     & -0.362    & -2.527      & -0.082      & \textbf{0.703}         \\
\midrule
$\alpha$ & 0.4     & 0.6       & 0.7       & 0.5         & 0.6         & 0.7           \\
$\beta$  & 0.6     & 0.4       & 0.5       & 0.6         & 0.6         & 0.5           \\
$\gamma$ & 0       & 0         & 0.2       & 0.1         & 0.2         & 0.2           \\
$\delta$   & 0.04    & 0.04      & 0.04      & 0.04        & 0.04        & 0.04          \\
\bottomrule            
	\end{tabular}
	\label{tab:exp}
\vspace{-1em}
\end{table}

\begin{figure*}[t]
	\centering
	\includegraphics[width=\linewidth]{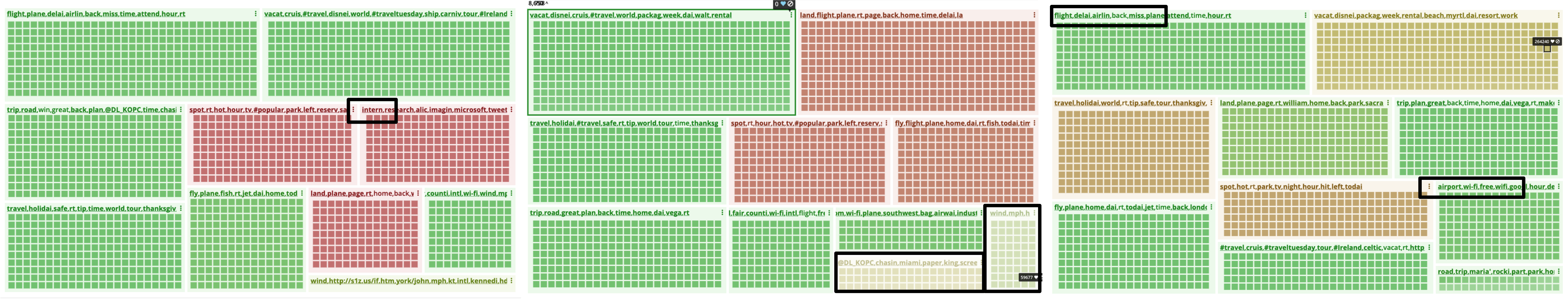}
\vspace{-1.8em}
	\caption{Exploring the Twitter dataset. After the initial iteration (left), some travel-related topics are found.
After adding ``intern'' to the stopword list, irrelevant topics are still included (middle). 
Tweets are further sifted (right).}
	\label{fig:tweet-s1}
\vspace{-1.8em}
\end{figure*}

\begin{figure}[t]
	\centering
	\includegraphics[width=0.49\linewidth]{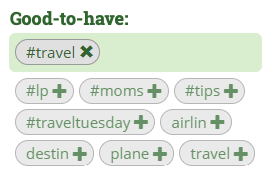}
	\includegraphics[width=0.49\linewidth]{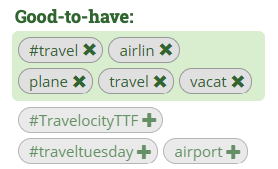}
\vspace{-0.5em}
	\caption{Initial user-input good-to-have keyword ``\#travel'' and the keywords recommended by TopicSifter (left).
The recommended keywords are also incorporated into the good-to-have list.}
	\label{fig:tweet-13}
\vspace{-1.5em}
\end{figure}

\subsection{Use Case 1: Exploring Scraped Data}
Jim is an art-major student who is also interested in technology. He is looking for technology areas where he can incorporate his artistic sense, and uses TopicSifter to retrieve talks related to his interest.

His visual exploration starts with an initial topic modeling that shows ten topics of all documents of the TED transcript dataset.
From the main view (shown in Fig. \ref{fig:ted-s1}(left)), he observes that a variety of topics are covered in the TED dataset, thus, he decides to focus on his interest, art and technology.
He adds the keyword compound ``art, technology'' to the good-to-have keyword list and run the TopicSifter.
He discovers some topics that are not interesting, such as biology/medicine or economy related ones, and downvotes them by clicking the topic cells.
During the process, he finds out that a keyword ``laughter'' was the third-most frequent word in the TED dataset as he sees the history view from the detail panel (shown in the upper-right of Fig.~\ref{fig:ted-s24}).
He reminds that the TED dataset is based on the scripts of the talks, and the keyword ``laughter'' is usually used to describe audience's reaction in scripts.
He adds the keyword to the list of stopwords so that it cannot influence the sifting process (Left of Fig.~\ref{fig:ted-s24}).
As he proceeds, he sees a topic with keywords ``market, africa, dollar'' and downvotes it since it looks unrelated.
Here, TopicSifter does not simply removes all document in the downvoted topic.
It still retrieves target-relevant documents that was in the downvoted topic, accomplishing high recall.
For example, documents titled ``The surprising seeds of a big data revolution in healthcare'', and ``Tim Brown urges designers to think big'' are both highly related to art and technology field, and were assigned to the ``money'' topic.
They were survived by TopicSifter by taking account into overall relevancy.
He finds ``comput, robot'' topic inteseting, inspects its documents in the table view interesting, and upvotes it (Fig.~\ref{fig:ted-s1}(middle)).
In this topic, Jim finds out interesting topics and corresponding documents that contain contents about 3-D printer or human-computer interaction.
Finally, he continues iterations until he is satisfied with his target documents about art and technology (Fig. \ref{fig:ted-s1}(right)).

\subsection{Use Case 2: Exploring Social Media Data}
Now, we will follow the case of a marketer in a travel agency, Pam, who uses the proposed technique to sort out consumers' interests in travel experiences. Pam starts by loading the twitter dataset and looking at the initial topics.

Since Twitter is a social media platform, many tweets are about everyday life and emotions.
For example, Pam sees that some topics include top keywords such as: ``rt'', ``home'', ``day'', and ``today''.
She adds the keyword ``\#travel'' to the good-to-have list to observe users' behavior using hashtag keywords about traveling on Twitter.
As recommended good-to-have words pop up around the selected keywords, she selects relevant keywords among them such as ``airline'', ``plane'', ``travel'' and ``vacation'' to see broader user interests about traveling (Fig. \ref{fig:tweet-13}).
After a single sifting phase, she observes that a red (and thus less relevant) topic includes the keyword ``intern'' (rectangle in Fig. \ref{fig:tweet-s1}(left)).
Many tweets included in it are comments about ``internship'' such as ``Why are like 80\% of the PokerRoad intern applicants from Canada? [...]''.
She finds it strange that a topic about internship is retrieved for travel related targets.
As it turns out, the word ``international'', which is relevant to the targets, is stemmed to ``intern'', so tweets about internships are incorrectly identified as relevant.
Pam adds ``intern'' to the stopword list to avoid this issue.
After one iteration, the ``intern'' topic is removed (Fig. \ref{fig:tweet-s1}(middle)).
She spots an unusual topic ``wind,mph''.
Tweets in this topic are mostly automatically generated from a weather bot twitter account such as ``HD: Light Rain and Breezy and 52 F at New York/John F. Kennedy [...]''.
Another topic ``@DL\_KOPC,chasin,miami'' contains  various spam messages such as advertisement for a trip to Miami.
She downvoted these two topics to remove additional spamming tweets (black rectangles in Fig. \ref{fig:tweet-s1}(middle)).
At the next iteration, there are many casual tweets such as ``Family, food, games, and football. That's Thanksgiving.'' or ``Just chatted w/ Jane Lindskold \& husband Jim here at the airport. Very cool people.''.
Pam continues exploration to find out more specific tweets that represent customers' interests related to traveling (shown in Fig. \ref{fig:tweet-s1}(right)).
One big travel-related concern is ``flight delay'' as shown in the top-right topic in Fig. \ref{fig:tweet-s1}(right).
Another interest is ``free Wi-Fi'' as shown in the bottom-right topic in Fig. \ref{fig:tweet-s1}(right).
She starts designing travel packages that includes free WiFi options and flight delay insurances.
The application helped her realize customer concerns and customize the agency's products.

\section{Discussion}
\label{sec:discussion}
\noindent Iterative methods are computational methods that update approximate solutions over iterations.
In general, iterative methods have some stopping criteria or stopping rules to terminate the methods, based on their objective functions or evaluation measures, e.g., when a score converges to a local minimum.
Likewise, many visual analytics systems that adopt interactive machine learning or optimization methods utilize some form of measures to evaluate their tasks and application.
These measures can be kept internally for monitoring; or can be shown to the users as charts (e.g., ~\cite{el2018specex}) or some form of visual encodings (e.g.,~\cite{kim2017pive}) to inform users about the status of the current iteration.
In our case, the relevancy scores of documents can be used as a measure.
Unfortunately, our iterative retrieval approach not only updates the solution (which is the retrieved set of relevant documents), but also updates the target by which we measure the relevance scores of the documents.
For this reason, comparing the relevance scores between iterations are meaningless if the target has been changed.
That is, a higher relevance score in an iteration does not necessarily mean a better solution than a lower relevance score in another iteration.
One na\"{\i}ve walkaround would be to compute the relevance scores of previously retrieved documents against the current target.
However, this walkaround requires the system to store all historical results and calculate the relevance scores again at every iteration, which is not practical.
Instead, for the TopicSifter prototype system, we decided to show retrieval status changes similar to membership changes in clustering.
As explained in Fig.~\ref{fig:history}, the history view in the detail panel shows changes in retrieved documents the over iterations in the stacked bar chart.
In addition, we use a colored triangle mark to indicate if a topic has changed much from the previous iteration as in Fig.~\ref{fig:topic_color}.
These kinds of visual cues can guide the users' decision on when to stop the iteration (e.g., limited change between iterations)

\section{Conclusion}
\label{sec:conclusion}
In the paper, we proposed a novel sifting technique to solve search space reduction problem interactively and iteratively.
Our technique combined interactive target building and targeted topic modeling to sift through document collections and retrieve relevant document as many as possible.
As a proof of concept, we built an interactive search space reduction system which offers tight integration between the visualization and the underlying algorithms.

\acknowledgments{
The authors wish to thank Ji Yeon Kim for her help with illustration for the paper. 
This work was supported in part by NSF grant OAC-1642410 and IIS-1750474. 
Any opinions, findings and conclusions or recommendations expressed in this material are those of the authors and do not necessarily reflect the views of funding agencies.
}

\bibliographystyle{abbrv-doi}

\bibliography{ms}

\begin{thebibliography}{10}

\bibitem{infosky02}
K.~Andrews, W.~Kienreich, V.~Sabol, J.~Becker, G.~Droschl, F.~Kappe,
  M.~Granitzer, P.~Auer, and K.~Tochtermann.
\newblock The info{S}ky visual explorer: Exploiting hierarchical structure and
  document similarities.
\newblock {\em Information Visualization}, 1(3-4):166--181, 2002. doi: {{%
10\hspace{.1pt}\discretionary{.}{%
}{.}\hspace{.4pt}1057\discretionary{/}{%
}{/}palgrave\hspace{.1pt}\discretionary{.}{%
}{.}\hspace{.4pt}ivs\hspace{.1pt}\discretionary{.}{%
}{.}\hspace{.4pt}9500023}}


\bibitem{blei2003latent}
D.~M. Blei, A.~Y. Ng, and M.~I. Jordan.
\newblock Latent dirichlet allocation.
\newblock {\em Journal of machine Learning research}, 3:993--1022, 2003.

\bibitem{facetatlas10}
N.~Cao, J.~Sun, Y.-R. Lin, D.~Gotz, S.~Liu, and H.~Qu.
\newblock Facet{A}tlas: Multifaceted visualization for rich text corpora.
\newblock {\em IEEE Transactions on Visualization and Computer Graphics},
  16(6):1172--1181, Nov 2010. doi: {{%
10\hspace{.1pt}\discretionary{.}{%
}{.}\hspace{.4pt}1109\discretionary{/}{%
}{/}TVCG\hspace{.1pt}\discretionary{.}{%
}{.}\hspace{.4pt}2010\hspace{.1pt}\discretionary{.}{%
}{.}\hspace{.4pt}154}}


\bibitem{cheng2010you}
Z.~Cheng, J.~Caverlee, and K.~Lee.
\newblock You are where you tweet: a content-based approach to geo-locating
  twitter users.
\newblock In {\em Proc. the ACM International Conference on Information and
  Knowledge Management}, pp. 759--768. ACM, 2010.

\bibitem{visirr}
J.~Choo, H.~Kim, E.~Clarkson, Z.~Liu, C.~Lee, F.~Li, H.~Lee, R.~Kannan, C.~D.
  Stolper, J.~Stasko, and H.~Park.
\newblock {VisIRR}: A visual analytics system for information retrieval and
  recommendation for large-scale document data.
\newblock {\em ACM Transactions on Knowledge Discovery from Data},
  12(1):8:1--8:20, Jan. 2018. doi: {{%
10\hspace{.1pt}\discretionary{.}{%
}{.}\hspace{.4pt}1145\discretionary{/}{%
}{/}3070616}}


\bibitem{utopian13}
J.~Choo, C.~Lee, C.~K. Reddy, and H.~Park.
\newblock {UTOPIAN}: User-driven topic modeling based on interactive
  nonnegative matrix factorization.
\newblock {\em IEEE Transactions on Visualization and Computer Graphics},
  19(12):1992--2001, Dec 2013. doi: {{%
10\hspace{.1pt}\discretionary{.}{%
}{.}\hspace{.4pt}1109\discretionary{/}{%
}{/}TVCG\hspace{.1pt}\discretionary{.}{%
}{.}\hspace{.4pt}2013\hspace{.1pt}\discretionary{.}{%
}{.}\hspace{.4pt}212}}


\bibitem{clarkson2009resultmaps}
E.~Clarkson, K.~Desai, and J.~Foley.
\newblock Resultmaps: Visualization for search interfaces.
\newblock {\em IEEE Transactions on Visualization and Computer Graphics},
  15(6):1057--1064, 2009.

\bibitem{roseriver14}
W.~Cui, S.~Liu, Z.~Wu, and H.~Wei.
\newblock How hierarchical topics evolve in large text corpora.
\newblock {\em IEEE Transactions on Visualization and Computer Graphics},
  20(12):2281--2290, Dec 2014. doi: {{%
10\hspace{.1pt}\discretionary{.}{%
}{.}\hspace{.4pt}1109\discretionary{/}{%
}{/}TVCG\hspace{.1pt}\discretionary{.}{%
}{.}\hspace{.4pt}2014\hspace{.1pt}\discretionary{.}{%
}{.}\hspace{.4pt}2346433}}


\bibitem{Drake2017}
B.~Drake, T.~Huang, A.~Beavers, R.~Du, and H.~Park.
\newblock Event detection based on nonnegative matrix factorization: Ceasefire
  violation, environmental, and malware events.
\newblock In D.~Nicholson, ed., {\em Advances in Human Factors in
  Cybersecurity}, pp. 158--169. Springer International Publishing, New York
  City, USA, 2018. doi: {{%
10\hspace{.1pt}\discretionary{.}{%
}{.}\hspace{.4pt}1007\discretionary{/}{%
}{/}978\discretionary{%
}{-}{-}3\discretionary{%
}{-}{-}319\discretionary{%
}{-}{-}60585\discretionary{%
}{-}{-}2\_16}}


\bibitem{Du2017}
R.~Du, D.~Kuang, B.~Drake, and H.~Park.
\newblock {DC-NMF}: Nonnegative matrix factorization based on
  divide-and-conquer for fast clustering and topic modeling.
\newblock {\em Journal of Global Optimization}, 68(4):777--798, Aug. 2017. doi:
  {{%
10\hspace{.1pt}\discretionary{.}{%
}{.}\hspace{.4pt}1007\discretionary{/}{%
}{/}s10898\discretionary{%
}{-}{-}017\discretionary{%
}{-}{-}0515\discretionary{%
}{-}{-}z}}


\bibitem{el2018specex}
M.~El-Assady, F.~Sperrle, O.~Deussen, D.~A. Keim, and C.~Collins.
\newblock {Visual Analytics for Topic Model Optimization based on
  User-Steerable Speculative Execution}.
\newblock {\em IEEE Transactions on Visualization and Computer Graphics}, 2018.
  doi: {{%
10\hspace{.1pt}\discretionary{.}{%
}{.}\hspace{.4pt}1109\discretionary{/}{%
}{/}TVCG\hspace{.1pt}\discretionary{.}{%
}{.}\hspace{.4pt}2018\hspace{.1pt}\discretionary{.}{%
}{.}\hspace{.4pt}2864769}}


\bibitem{fowler1991integrating}
R.~H. Fowler, W.~A. Fowler, and B.~A. Wilson.
\newblock Integrating query thesaurus, and documents through a common visual
  representation.
\newblock In {\em Proc. the International ACM SIGIR Conference on Research and
  Development in Information Retrieval}, pp. 142--151. ACM, 1991.

\bibitem{gomez2014similarity}
E.~Gomez-Nieto, F.~San~Roman, P.~Pagliosa, W.~Casaca, E.~S. Helou, M.~C.~F.
  de~Oliveira, and L.~G. Nonato.
\newblock Similarity preserving snippet-based visualization of web search
  results.
\newblock {\em IEEE Transactions on Visualization and Computer Graphics},
  20(3):457--470, 2014.

\bibitem{harman1988towards}
D.~Harman.
\newblock Towards interactive query expansion.
\newblock In {\em Proc. the International ACM SIGIR Conference on Research and
  Development in Information Retrieval}, pp. 321--331. ACM, 1988.

\bibitem{havre2001interactive}
S.~Havre, E.~Hetzler, K.~Perrine, E.~Jurrus, and N.~Miller.
\newblock Interactive visualization of multiple query results.
\newblock In {\em Proc. the IEEE Symposium on Information Visualization}, pp.
  105--. IEEE Computer Society, Washington, DC, USA, 2001.

\bibitem{hearst1995tilebars}
M.~A. Hearst.
\newblock {TileBars}: visualization of term distribution information in full
  text information access.
\newblock In {\em Proc. the SIGCHI Conference on Human Factors in Computing
  Systems}, pp. 59--66. ACM Press/Addison-Wesley Publishing Co., 1995.

\bibitem{HERR201728}
D.~Herr, Q.~Han, S.~Lohmann, and T.~Ertl.
\newblock Hierarchy-based projection of high-dimensional labeled data to reduce
  visual clutter.
\newblock {\em Computers \& Graphics}, 62:28 -- 40, 2017. doi: {{%
10\hspace{.1pt}\discretionary{.}{%
}{.}\hspace{.4pt}1016\discretionary{/}{%
}{/}j\hspace{.1pt}\discretionary{.}{%
}{.}\hspace{.4pt}cag\hspace{.1pt}\discretionary{.}{%
}{.}\hspace{.4pt}2016\hspace{.1pt}\discretionary{.}{%
}{.}\hspace{.4pt}12\hspace{.1pt}\discretionary{.}{%
}{.}\hspace{.4pt}004}}


\bibitem{inspire04}
E.~Hetzler and A.~Turner.
\newblock Analysis experiences using information visualization.
\newblock {\em IEEE Computer Graphics and Applications}, 24(5):22--26, Sept
  2004. doi: {{%
10\hspace{.1pt}\discretionary{.}{%
}{.}\hspace{.4pt}1109\discretionary{/}{%
}{/}MCG\hspace{.1pt}\discretionary{.}{%
}{.}\hspace{.4pt}2004\hspace{.1pt}\discretionary{.}{%
}{.}\hspace{.4pt}22}}


\bibitem{hoeber2006comparative}
O.~Hoeber and X.~D. Yang.
\newblock A comparative user study of web search interfaces: Hotmap, concept
  highlighter, and google.
\newblock In {\em Proc. the IEEE/WIC/ACM International Conference on Web
  Intelligence}, pp. 866--874. IEEE, 2006.

\bibitem{hoeber2005visualization}
O.~Hoeber, X.-D. Yang, and Y.~Yao.
\newblock Visualization support for interactive query refinement.
\newblock In {\em Proc. the IEEE/WIC/ACM International Conference on Web
  Intelligence}, pp. 657--665. IEEE, 2005.

\bibitem{convisit2015}
E.~Hoque and G.~Carenini.
\newblock Convisit: Interactive topic modeling for exploring asynchronous
  online conversations.
\newblock In {\em Proc. the International Conference on Intelligent User
  Interfaces}, pp. 169--180. ACM, New York, NY, USA, 2015. doi: {{%
10\hspace{.1pt}\discretionary{.}{%
}{.}\hspace{.4pt}1145\discretionary{/}{%
}{/}2678025\hspace{.1pt}\discretionary{.}{%
}{.}\hspace{.4pt}2701370}}


\bibitem{mcon2018}
E.~Hoque and G.~Carenini.
\newblock Interactive topic hierarchy revision for exploring a collection of
  online conversations.
\newblock {\em Information Visualization}, 0(0):1473871618757228, 2018. doi:
  {{%
10\hspace{.1pt}\discretionary{.}{%
}{.}\hspace{.4pt}1177\discretionary{/}{%
}{/}1473871618757228}}


\bibitem{kim2017pive}
H.~Kim, J.~Choo, C.~Lee, H.~Lee, C.~K. Reddy, and H.~Park.
\newblock {PIVE}: Per-iteration visualization environment for real-time
  interactions with dimension reduction and clustering.
\newblock In {\em Proc. the AAAI Conference on Artificial Intelligence}, 2017.

\bibitem{topiclens17}
M.~Kim, K.~Kang, D.~Park, J.~Choo, and N.~Elmqvist.
\newblock {TopicLens}: Efficient multi-level visual topic exploration of
  large-scale document collections.
\newblock {\em IEEE Transactions on Visualization and Computer Graphics},
  23(1):151--160, Jan 2017. doi: {{%
10\hspace{.1pt}\discretionary{.}{%
}{.}\hspace{.4pt}1109\discretionary{/}{%
}{/}TVCG\hspace{.1pt}\discretionary{.}{%
}{.}\hspace{.4pt}2016\hspace{.1pt}\discretionary{.}{%
}{.}\hspace{.4pt}2598445}}


\bibitem{kdd2013}
D.~Kuang and H.~Park.
\newblock Fast rank-2 nonnegative matrix factorization for hierarchical
  document clustering.
\newblock In {\em Proc. the {ACM} {SIGKDD} International Conference on
  Knowledge Discovery and Data Mining}, pp. 739--747. {ACM}, 2013. doi: {{%
10\hspace{.1pt}\discretionary{.}{%
}{.}\hspace{.4pt}1145\discretionary{/}{%
}{/}2487575\hspace{.1pt}\discretionary{.}{%
}{.}\hspace{.4pt}2487606}}


\bibitem{jogo2015}
D.~Kuang, S.~Yun, and H.~Park.
\newblock {SymNMF}: nonnegative low-rank approximation of a similarity matrix
  for graph clustering.
\newblock {\em Journal of Global Optimization}, 62(3):545--574, Nov. 2014. doi:
  {{%
10\hspace{.1pt}\discretionary{.}{%
}{.}\hspace{.4pt}1007\discretionary{/}{%
}{/}s10898\discretionary{%
}{-}{-}014\discretionary{%
}{-}{-}0247\discretionary{%
}{-}{-}2}}


\bibitem{ivisclustering12}
H.~Lee, J.~Kihm, J.~Choo, J.~Stasko, and H.~Park.
\newblock {iVisClustering}: An interactive visual document clustering via topic
  modeling.
\newblock {\em Computer Graphics Forum}, 31(3pt3):1155--1164, 2012. doi: {{%
10\hspace{.1pt}\discretionary{.}{%
}{.}\hspace{.4pt}1111\discretionary{/}{%
}{/}j\hspace{.1pt}\discretionary{.}{%
}{.}\hspace{.4pt}1467\discretionary{%
}{-}{-}8659\hspace{.1pt}\discretionary{.}{%
}{.}\hspace{.4pt}2012\hspace{.1pt}\discretionary{.}{%
}{.}\hspace{.4pt}03108\hspace{.1pt}\discretionary{.}{%
}{.}\hspace{.4pt}x}}


\bibitem{levy2014neural}
O.~Levy and Y.~Goldberg.
\newblock Neural word embedding as implicit matrix factorization.
\newblock In {\em Proc. the International Conference on Neural Information
  Processing Systems - Volume 2}, pp. 2177--2185. MIT Press, Cambridge, MA,
  USA, 2014.

\bibitem{li2014req}
C.~Li, Y.~Wang, P.~Resnick, and Q.~Mei.
\newblock Req-rec: High recall retrieval with query pooling and interactive
  classification.
\newblock In {\em Proc. the International ACM SIGIR Conference on Research and
  Development in Information Retrieval}, pp. 163--172. ACM, 2014.

\bibitem{magdy2010pres}
W.~Magdy and G.~J. Jones.
\newblock {PRES}: a score metric for evaluating recall-oriented information
  retrieval applications.
\newblock In {\em Proc. the International ACM SIGIR Conference on Research and
  Development in Information Retrieval}, pp. 611--618. ACM, 2010.

\bibitem{manning2010introduction}
C.~Manning, P.~Raghavan, and H.~Sch{\"u}tze.
\newblock Introduction to information retrieval.
\newblock {\em Natural Language Engineering}, 16(1):100--103, 2010.

\bibitem{mikolov2013efficient}
T.~Mikolov, K.~Chen, G.~Corrado, and J.~Dean.
\newblock Efficient estimation of word representations in vector space.
\newblock {\em arXiv preprint arXiv:1301.3781}, 2013.

\bibitem{forcespire12}
C.~North, A.~Endert, and P.~Fiaux.
\newblock Semantic interaction for sensemaking: Inferring analytical reasoning
  for model steering.
\newblock {\em IEEE Transactions on Visualization and Computer Graphics},
  18:2879--2888, 2012. doi: {{%
doi\hspace{.1pt}\discretionary{.}{%
}{.}\hspace{.4pt}ieeecomputersociety\hspace{.1pt}\discretionary{.}{%
}{.}\hspace{.4pt}org\discretionary{/}{%
}{/}10\hspace{.1pt}\discretionary{.}{%
}{.}\hspace{.4pt}1109\discretionary{/}{%
}{/}TVCG\hspace{.1pt}\discretionary{.}{%
}{.}\hspace{.4pt}2012\hspace{.1pt}\discretionary{.}{%
}{.}\hspace{.4pt}260}}


\bibitem{pennington2014glove}
J.~Pennington, R.~Socher, and C.~Manning.
\newblock Glove: Global vectors for word representation.
\newblock In {\em Proc. the Conference on Empirical Methods in Natural Language
  Processing}, pp. 1532--1543, 2014.

\bibitem{porter1980algorithm}
M.~F. Porter.
\newblock An algorithm for suffix stripping.
\newblock {\em Program}, 14(3):130--137, 1980.

\bibitem{rakesh2018sparse}
V.~Rakesh, W.~Ding, A.~Ahuja, N.~Rao, Y.~Sun, and C.~K. Reddy.
\newblock A sparse topic model for extracting aspect-specific summaries from
  online reviews.
\newblock In {\em Proc. the World Wide Web Conference}, pp. 1573--1582.
  International World Wide Web Conferences Steering Committee, 2018.

\bibitem{Reiterer2005}
H.~Reiterer, G.~Tullius, and T.~M. Mann.
\newblock Insyder: a content-based visual-information-seeking system for the
  web.
\newblock {\em International Journal on Digital Libraries}, 5(1):25--41, Mar
  2005. doi: {{%
10\hspace{.1pt}\discretionary{.}{%
}{.}\hspace{.4pt}1007\discretionary{/}{%
}{/}s00799\discretionary{%
}{-}{-}004\discretionary{%
}{-}{-}0111\discretionary{%
}{-}{-}y}}


\bibitem{ruotsalo2015interactive}
T.~Ruotsalo, G.~Jacucci, P.~Myllym{\"a}ki, and S.~Kaski.
\newblock Interactive intent modeling: information discovery beyond search.
\newblock {\em Communications of the ACM}, 58(1):86--92, 2015.

\bibitem{ruotsalo2013directing}
T.~Ruotsalo, J.~Peltonen, M.~Eugster, D.~G{\l}owacka, K.~Konyushkova,
  K.~Athukorala, I.~Kosunen, A.~Reijonen, P.~Myllym{\"a}ki, G.~Jacucci, et~al.
\newblock Directing exploratory search with interactive intent modeling.
\newblock In {\em Proc. the ACM International Conference on Information and
  Knowledge Management}, pp. 1759--1764. ACM, 2013.

\bibitem{ruthven2003survey}
I.~Ruthven and M.~Lalmas.
\newblock A survey on the use of relevance feedback for information access
  systems.
\newblock {\em The Knowledge Engineering Review}, 18(2):95--145, 2003.

\bibitem{hltm2018}
A.~Smith, V.~Kumar, J.~Boyd-Graber, K.~Seppi, and L.~Findlater.
\newblock Closing the loop: User-centered design and evaluation of a
  human-in-the-loop topic modeling system.
\newblock In {\em Proc. the International Conference on Intelligent User
  Interfaces}, pp. 293--304. ACM, New York, NY, USA, 2018. doi: {{%
10\hspace{.1pt}\discretionary{.}{%
}{.}\hspace{.4pt}1145\discretionary{/}{%
}{/}3172944\hspace{.1pt}\discretionary{.}{%
}{.}\hspace{.4pt}3172965}}


\bibitem{smith2006facetmap}
G.~Smith, M.~Czerwinski, B.~Meyers, D.~Robbins, G.~Robertson, and D.~S. Tan.
\newblock {FacetMap}: A scalable search and browse visualization.
\newblock {\em IEEE Transactions on Visualization and Computer Graphics},
  12(5):797--804, 2006.

\bibitem{wang2016targeted}
S.~Wang, Z.~Chen, G.~Fei, B.~Liu, and S.~Emery.
\newblock Targeted topic modeling for focused analysis.
\newblock In {\em Proc. the ACM SIGKDD International Conference on Knowledge
  Discovery and Data Mining}, pp. 1235--1244. ACM, 2016.

\bibitem{wang2018disentangling}
S.~Wang, M.~Zhou, S.~Mazumder, B.~Liu, and Y.~Chang.
\newblock Disentangling aspect and opinion words in target-based sentiment
  analysis using lifelong learning.
\newblock {\em arXiv preprint arXiv:1802.05818}, 2018.

\bibitem{white2009exploratory}
R.~{White} and R.~{Roth}.
\newblock {\em Exploratory Search: Beyond the Query-Response Paradigm}.
\newblock Morgan \& Claypool, 2013.

\bibitem{zheng2017semi}
X.~Zheng, A.~Sun, S.~Wang, and J.~Han.
\newblock Semi-supervised event-related tweet identification with dynamic
  keyword generation.
\newblock In {\em Proc. of the ACM on Conference on Information and Knowledge
  Management}, pp. 1619--1628. ACM, 2017.

\end{thebibliography}
\end{document}